%% file: article.tex
\documentclass[preprint, 5p]{elsarticle}
\title{In operando active learning of interatomic interaction during large-scale simulations}
\author[]{M.~Hodapp\fnref{fn1}\corref{cor1}}
\author[]{A.~Shapeev\fnref{fn2}}
\fntext[fn1]{\textit{E-mail address:} m.hodapp@skoltech.ru}
\fntext[fn2]{\textit{E-mail address:} a.shapeev@skoltech.ru}
\cortext[cor1]{Corresponding author}
\address{\vspace{0.5em}Skolkovo Institute of Science and Technology (Skoltech), Center for Energy Science and Technology, Moscow (RU)}
\journal{Machine Learning: Science and Technology}
\usepackage{amssymb}
\usepackage{mathrsfs}
\usepackage{amsthm}
\usepackage[nonumberlist,nopostdot,sort=def,nowarn,acronym]{glossaries} 
\usepackage{glossary-mcols} 
\usepackage{color,upgreek}
\usepackage[ruled,lined]{algorithm2e}
\usepackage{setspace}
\usepackage{pifont}
\usepackage{array}
\usepackage{multirow}
\usepackage{multicol}
\usepackage{arydshln}
\usepackage{titlesec}
\usepackage{appendix}
\usepackage{fancybox,framed} 
\usepackage{manfnt} 
\usepackage{scrextend}
\usepackage{float}
\usepackage{enumitem}
\usepackage{mathtools} 
\usepackage{textcomp} 
\usepackage[bottom]{footmisc}

\usepackage{hyperref}
\input{style}

\makeglossaries

\glsdisablehyper 
\renewcommand{\glossarysection}[2][]{} 
\glssetwidest[0]{\hspace{3.5em}}

\newproof{prf}{Proof}

\newtheorem*{rem2*}{Important remark}

\renewcommand{\arraystretch}{1.4}

\titleformat*{\section}{\bfseries}

\makeatletter
\def\blfootnote{\gdef\@thefnmark{}\@footnotetext}
\makeatother

\newcommand{\latConst}{\ensuremath{a_0}}
\newcommand{\displ}{\ensuremath{u}}
\newcommand{\bdispl}{\ensuremath{\sty{\displ}}}
\newcommand{\bdisplDD}{\ensuremath{\tilde{\bdispl}}}
\newcommand{\Etot}{\ensuremath{\Pi}}
\newcommand{\Esite}{\ensuremath{\clE}}
\newcommand{\force}{\ensuremath{f}}
\newcommand{\bforce}{\ensuremath{\sty{\force}}}
\newcommand{\stressShear}{\ensuremath{\tau}}
\newcommand{\burgers}{\ensuremath{b}}
\newcommand{\bburgers}{\ensuremath{\sty{\burgers}}}
\newcommand{\atom}{\ensuremath{r}}
\newcommand{\Atom}{\ensuremath{\sty{\atom}}}
\newcommand{\Atoms}{\ensuremath{\{\Atom_i\}}}
\newcommand{\AtomsNew}{\ensuremath{\{\Atom_i^\ast\}}}
\newcommand{\Neigh}{\ensuremath{\{\Atom_{ij}\}}}
\newcommand{\NeighNew}{\ensuremath{\{\Atom_{ij}^\ast\}}}

\DeclareMathOperator{\arctantwo}{arctan2}

\begin{document}

\begin{frontmatter}
 \begin{abstract}
  A well-known drawback of state-of-the-art machine-learning interatomic potentials is their poor ability to extrapolate beyond the training domain.
  For small-scale problems with tens to hundreds of atoms this can be solved by using active learning which is able to select atomic configurations on which a potential attempts extrapolation and add them to the ab initio-computed training set.
  In this sense an active learning algorithm can be viewed as an on-the-fly interpolation of an ab initio model.
  For large-scale problems, possibly involving tens of thousands of atoms, this is not feasible because one cannot afford even a single density functional theory computation with such a large number of atoms.
  
  This work marks a new milestone toward fully automatic ab initio-accurate large-scale atomistic simulations.
  We develop an active learning algorithm that identifies local subregions of the simulation region where the potential extrapolates.
  Then the algorithm constructs periodic configurations out of these local, non-periodic subregions, sufficiently small to be computable with plane-wave density functional theory codes, in order to obtain accurate ab initio energies.
  We benchmark our algorithm on the problem of screw dislocation motion in bcc tungsten and show that our algorithm reaches ab initio accuracy, down to typical magnitudes of numerical noise in DFT codes.
  We show that our algorithm reproduces material properties such as core structure, Peierls barrier, and Peierls stress.
  This unleashes new capabilities for computational materials science toward applications which have currently been out of scope if approached solely by ab initio methods.
 \end{abstract}
\end{frontmatter}

\section{Introduction}
\label{sec:intro}

It is now understood that the design of novel materials with exotic, unprecedented mechanical properties, specifically metallic alloys, requires a computer-assisted approach since many of the underlying mechanisms cannot be directly quantified by real experiments and are often not even observed. The state-of-the-art computing hardware and algorithms can simulate hundreds of atoms by direct atomistic simulations with density functional theory (DFT) which allows for an accurate prediction of ideal crystalline properties, such as phase stability or elastic moduli. However, on the macroscopic scale many critical material properties, such as yield strength or fracture toughness, are determined by the microstructure and, hence, require a multiscale approach.
Presently, multiscale simulations are prohibitively expensive with DFT and, in lieu thereof, carried out with empirical interatomic potentials. However, existing empirical potentials can, at their best, only suggest a qualitative mechanism but remain inadequate for a quantitative understanding of multicomponent systems and chemical interactions.

A popular multiscale approach in the materials science community are quantum mechanics/molecular mechanics \citep[QM/MM,][]{ogata_hybrid_2001,woodward_flexible_2002,csanyi_learn_2004,choly_multiscale_2005,bernstein_hybrid_2009} methods, where only a small subset of the computational domain, usually in the vicinity of crystalline defects, is treated quantum-mechanically while the remainder is approximated by interatomic potentials. A newer class of methods are classical atomistic simulations using machine-learning interatomic potentials \citep[MLIPs, e.g.,][]{behler_perspective:_2016} which---in comparison with empirical potentials---admit a flexible and generic functional form allowing to solve any problem with arbitrary accuracy, at least in theory (cf. \citep{shapeev_moment_2016}). This makes them a promising candidate for multiscale simulations since using an interatomic potential everywhere is orders of magnitudes faster than retaining a quantum mechanical model in parts of the computational domain.

Based on these premisses, different MLIPs have been proposed in recent years, mainly differing in their representation.
Recent review papers \citep{nyshadham_machine-learned_2019,zuo_performance_2019} compare a number of interatomic potentials, namely the neural network potential \citep[NNP,][]{behler_generalized_2007}, the Gaussian approximation potential \citep[GAP,][]{bartok_gaussian_2010,bartok_representing_2013}, the spectral neighbor analysis potential \citep[SNAP,][]{thompson2015-SNAP} and the moment tensor potential \citep[MTP,][]{shapeev_moment_2016}. Differences in the potential representation mainly influence the computational cost but all the above-mentioned MLIPs are able to produce (qualitatively) similar results.
Several other representations and their variants have been proposed to date \cite{botu2015-MLIP,schutt2017-schnet,jinnouchi2019-on-the-fly,vandermause2020-bayesian-on-the-fly,drautz2019-ace,smith2017-ANI,vanderoord2020-pip}, including related approaches in molecular modeling \cite{christensen2020-fchl-revisited}.
Hence, MLIPs can now be considered as sufficiently mature and, consequently, applications are rapidly emerging.
The first notable contributions toward applications in computational metallurgy are the GAPs for bcc tungsten and iron by \citet{szlachta_first_2014,dragoni_energetics_2016}, and the Al-Mg-Si NNP by \citet{kobayashi_neural_2017}. It was shown by the authors that these potentials can reproduce crucial properties that govern the mechanical and thermodynamical behavior, e.g., vacancy formation energies, stacking fault energies or phonon spectra. 

On the other hand, a major drawback of state-of-the-art MLIPs is their poor extrapolation beyond the known training configurations. Therefore, the most time-consuming step in developing these MLIPs is the construction of the training set. It takes months, or sometimes years, to manually construct all the needed configurations that would be representative of all configurations appearing in a simulation; this is done based on physical intuition and best practices (see the PhD theses \citep{szlachta_accuracy_2014,dragoni_achieving_2018}). This approach might work well for simple problems, e.g., single-component materials, in which atomistic mechanisms of many phenomena are well-understood. However, a manual, ``human-developed'' procedure will be too involved to effectively address more complex scenarios. For example, the GAP for iron \citep{dragoni_achieving_2018} has indeed been shown to successfully reproduce crucial properties of screw \citep{maresca_screw_2018} but \emph{not} of edge dislocations \citep{fellinger_geometries_2018}.

A different approach is \emph{active learning} \citep{settles_active_2010}. In contrast to passive learning, where the training set is finalized before running a simulation, any configuration \emph{appearing} in a simulation is allowed to become a part of the training set. In a nutshell, the idea is to estimate the magnitude of the error of the MLIP, its \emph{extrapolation grade}, during the simulation and, should it exceed some threshold, recompute energies and forces (and stresses, if necessary) with the underlying ab initio model and retrain the potential including the newly acquired data. In the field of atomistic modeling, active learning has been initially proposed in the context of QM/MM methods by \citet{de_vita_novel_1997,csanyi_learn_2004}---to learn force fields of recurring atomic neighborhoods on-the-fly during the simulation---and continues to develop to date \citep{li_molecular_2015,bianchini_enabling_2019,jinnouchi2019-on-the-fly}. For MLIPs, an active learning method has been proposed by \citet{behler_representing_2014} using a query-by-committee strategy (cf. \citep[][]{settles_active_2010}). Another method based on D-optimality was proposed by \citet{podryabinkin_active_2017} who estimate the extrapolation grade by means of the growth in the determinant of the system matrix associated with the loss function. Their method has been successfully applied to predict molecular properties \citep{gubaev_machine_2018}, diffusion processes in metals \citep{novoselov_moment_2018} and phase stability of (random) alloys \citep{gubaev_accelerating_2019,podryabinkin_accelerating_2019}, alongside with a reduction of the necessary amount of DFT calculations by several orders of magnitude.
However, despite these recent successes, applying active learning to larger simulations with thousand or more atoms requires a novel approach since DFT codes cannot efficiently handle thousands of atoms.
Such ``larger simulations'' are necessary to solve problems which involve extended defects with important properties affecting the material microstructure, e.g., dislocations, cracks or grain boundaries;
typical properties are their mobility, possibly altered by interactions of the defect with impurities (or other types of defects), and their long-range behavior.
In particular, the number of atoms rapidly exceeds the tens of thousands when considering intrinsically three-dimensional extended defects, such as kinked or jogged dislocations.

Building on \citep{podryabinkin_active_2017}, the main aim of the present work is thus to develop an active learning algorithm for large-scale problems containing extended defects, such as dislocations. The idea we pursue is as follows: from the large-scale computational domain we extract local configurations of 100--200 atoms, small enough to be feasible to compute with DFT, and measure the extrapolation grade therein. Should the extrapolation grade in one of them become too large, we automatically convert it to a periodic configuration maintaining the atomic neighborhoods from the large-scale problem, run a DFT calculation on this configuration and subsequently extend the training set.

While the algorithm, described in the previous paragraph, appears similar to the ``learning-on-the-fly QM/MM'', it is \emph{novel in our understanding} and does not share some of the drawbacks of QM/MM as will be made clear in the following. The distinguishing feature of our algorithm is the use of a MLIP in the \emph{entire} computational domain---as opposed to the direct coupling between DFT and interatomic potentials in QM/MM---which allows us to compute energetic quantities \emph{directly}, such as energy/activation barriers or defect formation energies. In QM/MM, energetic quantities are usually computed using indirect methods (e.g., \citep{swinburne_computing_2017}) since the quantum mechanical energy is a global quantity and cannot be computed for clusters of atoms.
Nevertheless, it is still necessary that the periodic training configurations are not ``too'' artificial to avoid degrading the accuracy of the MLIP. In Section \ref{sec:per_cfgs} we will present a new method to construct such periodic training configurations by symmetrizing the atomic positions so that atoms at the domain boundaries enjoy a neighborhood close those in the large-scale problem.

We test our methodology by predicting properties characterizing screw dislocation motion in bcc tungsten, i.e., we perform simulations in order to calculate the core structure, Peierls barrier and Peierls stress. Thereby, the focus is set on demonstrating that our training procedure ensures that the MLIP is able to (i) learn the \emph{right} training configurations during these simulations and (ii) reliably predicts energy differences and forces. 


\section{Active learning algorithm}
\label{sec:algo}

Let $\Atoms$ be a configuration of atoms subject to general (usually non-periodic) boundary conditions. We assume that the atomistic model is local in the sense that the motion of each atom $\Atom_i$ depends on its relative position $\Atom_{ij} = \Atom_j - \Atom_i$ with respect to all other atoms $\Atom_j$ within its neighborhood which usually extends to a few lattice spacings, i.e., up to third- or fourth nearest neighbors. The total energy $\Etot$ then admits a partitioning into per-atom energies $\Esite = \Esite(\Neigh)$ leading to
\begin{equation}
 \Etot = \sum_i \Esite(\Neigh).
\end{equation}
In this work we model the per-atom energies using moment tensor potentials \citep[MTP,][]{shapeev_moment_2016} such that
\begin{equation}\label{eq:Eatom}
 \Esite(\Neigh) = \Esite(\Neigh; \umtheta) = \sum_\alpha \mtheta_\alpha B_\alpha(\Neigh),
\end{equation}
where the $\mtheta_\alpha$'s are the fitting parameters and $B_\alpha(\Neigh)$ the basis functions. For further details on the potential representation we refer to \ref{sec:mtp_rep}.

In order to judge on the accuracy of the MTP, there exist a number of active learning algorithms \cite{behler_representing_2014,zhang2019-e-active-learning,smith2018-active-learning,jinnouchi2019-on-the-fly,vandermause2020-bayesian-on-the-fly} that are able to automatically assess whether $\Atoms$ is well-represented in the training set.
If not, one could add $\Atoms$ to the training set after computing its DFT energy and gradients.
However, if $\Atoms$ consists of 1000 or more atoms then this strategy is not feasible: DFT is very time-consuming and scales cubically with the total number of atoms.

In this work we develop an algorithm to learn interatomic interaction, as given by DFT, by automatically extracting small, representative configurations of 100--200 atoms with active learning and composing our training set from these small configurations.
Our algorithm has two components:
\begin{itemize}
 \item[(i)] detection of extrapolative neighborhoods,
 \item[(ii)] completion of such neighborhoods to small periodic configurations efficiently computable with plane-wave DFT codes.
\end{itemize}
Our algorithm of extracting neighborhoods is based on the D-optimality criterion \citep{settles_active_2010} developed for MTPs in \cite{podryabinkin_active_2017,gubaev_accelerating_2019}.
In particular, we adapt the algorithm for detecting per-neighborhood extrapolation.
In such an algorithm we determine the extrapolation grade $\gamma = \gamma(\NeighNew)$ for a neighborhood $\NeighNew$ of the $i$-th atom in a configuration based on how far the descriptors, $B_\alpha(\NeighNew)$, are from those of the training set.
That is, we imagine the existence of per-atom energies $\Esite^{\rm qm}$ and fitting the parameters $\umtheta$ to them:
\begin{equation}\label{eq:fitting}
 \Esite(\Neigh; \umtheta) - \Esite^{\rm qm}(\Neigh) = 0.
\end{equation}
Since \eqref{eq:fitting} would be overdetermined, we select an $m \times m$ submatrix $\uuA$ of the system matrix associated with \eqref{eq:fitting}. The $m$ rows in $\uuA$, representing a neighborhood $\Neigh$ from the training set, are chosen to maximize $\abs{\det{\uuA}}$ using the maxvol algorithm \citep{olshevsky_how_2010}. The extrapolation grade is then computed by solving the linear system (cf. \citep{podryabinkin_active_2017}, Section 3.3)
\begin{equation}\label{eq:comp_extrap_grade}
 \begin{aligned}
  \uc &= \ub^\sT\uuA^{-1} \\
  &=
  \begin{pmatrix}
   B_1(\NeighNew) & \cdots & B_m(\NeighNew)
  \end{pmatrix}
  \begin{pmatrix}
   B_1(\Neigh_1) & \cdots & B_m(\Neigh_1) \\
   \vdots & \ddots & \vdots \\
   B_1(\Neigh_m) & \cdots & B_m(\Neigh_m)
  \end{pmatrix}^{-1}, \\[1em]
  &\text{such that} \; \gamma = \underset{i}{\max\,}{\abs{c_i}}.
 \end{aligned}
\end{equation}
To be more precise, we calculate the extrapolation grade for nonlinearly parameterized potentials; this calculation is a natural generalization of the linear case (the one with basis functions $B_\alpha(\Neigh)$) developed in \cite{gubaev_accelerating_2019} and outlined in \ref{sec:mtp_nln}.
This grade allows us to formulate a practically universal criterion by using a threshold for permissible extrapolation.

\begin{figure}[hbt]
 \centering
 \includegraphics[width=0.9\textwidth]{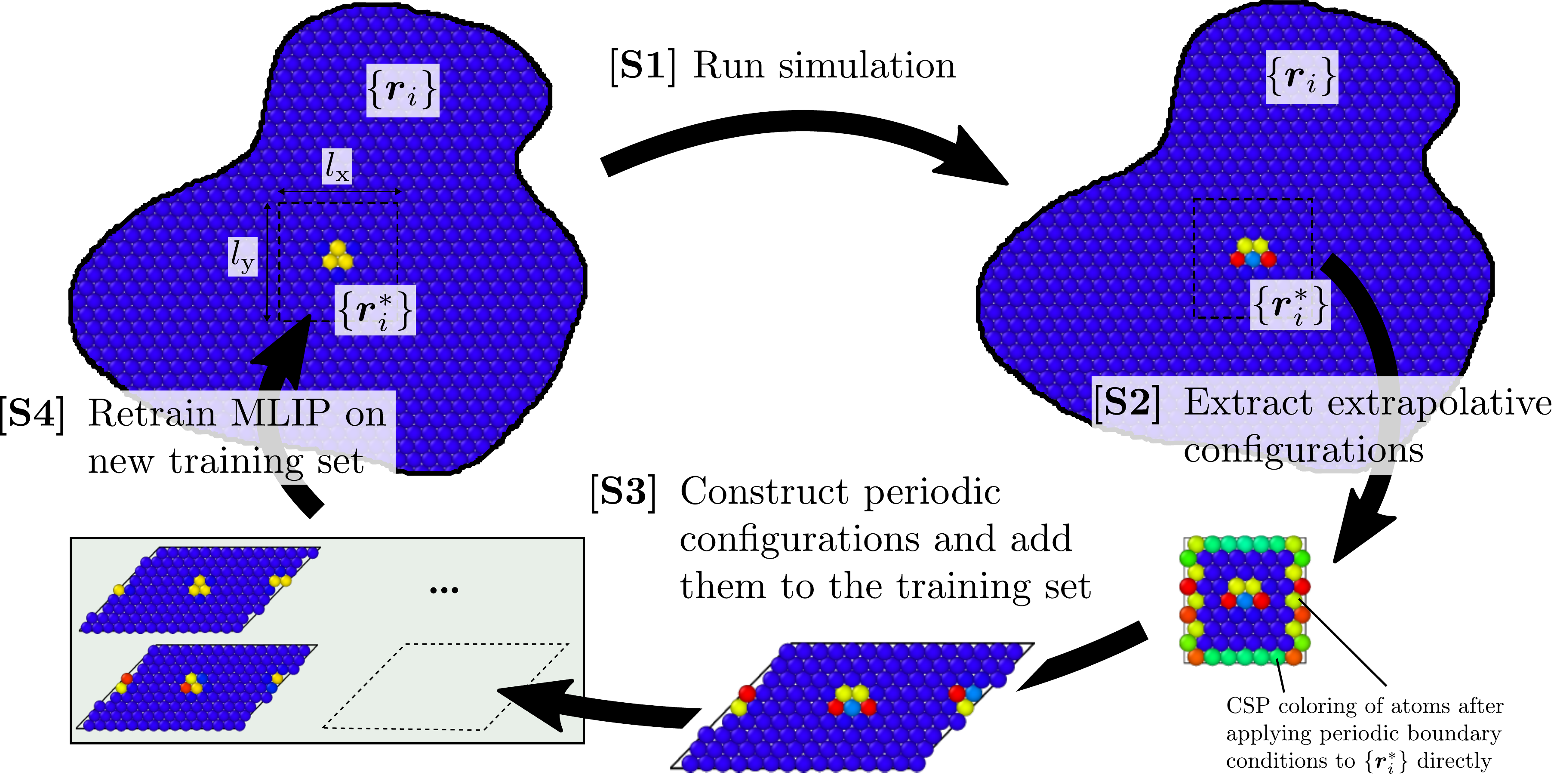}
 \caption{Schematic illustration of the individual steps during an atomistic simulation for screw dislocation motion using the active learning algorithm. \textbf{[S1]} The simulation runs until active learning detects inadmissible extrapolation in $\AtomsNew$ containing the dislocation core atoms (assuming that the MLIP has been already trained sufficiently well on the ideal crystal structure). \textbf{[S2]} Extraction of the extrapolative $\AtomsNew$. \textbf{[S3]} Construction of periodic configurations which capture the extrapolative neighborhoods of $\AtomsNew$ but removes the artificial ones at the boundary in order to accurately compute the dislocation core energy. \textbf{[S4]} The potential is retrained on the updated training set. The coloring of the atoms is due to the centrosymmetry parameter \citep[CSP,][]{kelchner_dislocation_1998} in order visualize their deviation from the bulk crystal configuration}
 \label{fig:al_algo_schematic}
\end{figure}

If the size of our atomistic configuration $\Atoms$ was $\approx\,$100--200 atoms, then our job would be done---we add this configuration to the training set together with the DFT-computed energy, forces, and stresses and continue (or restart) the simulation. Again, this is not feasible with larger systems because collecting DFT data scales cubically with the system size. Nevertheless, if the extrapolative neighborhoods $\NeighNew$ appear in \emph{subsets} $\AtomsNew \subseteq \Atoms$ of 100--200 atoms which are sufficiently far from each other, we can construct \emph{several} training configurations which, together, indeed capture all extrapolative neighborhoods. The justification of this approach is evident if the boundary atoms of a configuration $\AtomsNew$ have neighborhoods which are close to the ones they would have in the corresponding training configuration obtained by periodically replicating $\AtomsNew$. This is of course not the case if $\AtomsNew$ contains extended defects, such as dislocations, as illustrated in Figure \ref{fig:al_algo_schematic}. Therefore, running a DFT calculation on $\AtomsNew$ after naively applying periodic boundary conditions would critically degrade the accuracy of the MTP by training it on many artificial neighborhoods that occur at the boundary (guide the eye to the coloring of the atoms).

Our novel contribution here is the extension of active learning to precisely such cases by combining the extrapolation detection with the second component, the construction of periodic training configurations which capture the extrapolative neighborhoods $\NeighNew$ appearing in $\AtomsNew$. From Figure \ref{fig:al_algo_schematic} it can be seen that this is a more involved endeavour since the number of atoms as well as the shape of this periodic configuration differ from $\AtomsNew$. We postpone the details of this construction to the next section and first describe the full active learning algorithm.

During an atomistic simulation we may frequently encounter extrapolative configurations. Instead of retraining the potential immediately after detecting any extrapolation, we propose a greedy strategy that runs the simulation for a certain number of iterations\footnote{Here, iteration either refers to a step of a nonlinear solver, in the case of structural relaxation, or a molecular dynamics timestep} and subsequently selects a minimal set of configurations which \emph{must be} added to the training set in order to ensure that $\gamma$ remains below some threshold. In this way we allow for training configurations further outside the training set which potentially interpolate on otherwise selected configurations using an iteration-wise scheme. The individual steps of the full algorithm, which is schematically depicted in Figure \ref{fig:al_algo_schematic}, are described below:

\bigskip
\doublebox{
 \begin{minipage}{0.9\textwidth}
  \begin{itemize}
   \item[\textbf{[S0]}]
   Define the atomistic configuration $\Atoms$ and select the subset(s) $\AtomsNew\subseteq\Atoms$ where to compute the extrapolation grades of the atomic neighborhoods.
   \item[\textbf{[S1]}]
   Run the atomistic simulation for $N$ iterations. For all $n = 1,...,N$, compute the highest extrapolation grade $\gamma_n$ in $\AtomsNew_n$ after every iteration $n$.
   \item[\textbf{[S2]}]
   Stop the simulation after the $N$ iterations---or if $\gamma_n$ exceeds a chosen threshold $\gamma_{\rm max}$---and add all configurations $\AtomsNew_n$ whose extrapolation grade is larger than $\gamma_{\rm min}$ to the set of \emph{training candidates}.
   \item[\textbf{[S3]}]
   Update the training set using the following query strategy:
   \begin{itemize}[leftmargin=1.23cm]
    \item[\textbf{[S3.0]}]
    Convert all configurations $\AtomsNew$ in the set of training candidates to \emph{periodic configurations} according to the steps (1)--(4) illustrated in Figure \ref{fig:construct_dft_cells}.
    \item[\textbf{[S3.1]}]
    Move the periodic configuration with the highest extrapolation grade from the training candidates to the \emph{training set}.
    \item[\textbf{[S3.2]}]
    Update the matrix $\uuA$ from \eqref{eq:comp_extrap_grade} using the maxvol algorithm and recompute the extrapolation grades for all remaining training candidates.
    \item[\textbf{[S3.3]}]
    Go to \textbf{[S4]} if the extrapolation grades for all remaining training candidates are $\le \gamma_{\rm min}$ or if no training candidates are left. Otherwise go back to \textbf{[S3.1]}.
   \end{itemize}
   \item[\textbf{[S4]}]
   Retrain the potential and go back to \textbf{[S1]} until $\forall\, n=1,...,N \; \gamma_n \le \gamma_{\rm min}$.
  \end{itemize}
 \end{minipage}
}
\bigskip

\noindent%
The steps \textbf{[S1]}--\textbf{[S4]} are then repeated until convergence or the maximum number of iterations is attained.

\section{Construction of periodic training configurations}
\label{sec:per_cfgs}

\subsection{Existing methods within the scope of QM/MM}

We now turn to the question of how to construct an appropriate training set from configurations $\AtomsNew$ in which the extrapolation grade exceeds a certain threshold. One of the currently most accurate and efficient ways to compute quantum mechanical quantities, such as energies, forces and stresses, appears to be plane-wave, i.e., periodic, density functional theory (DFT) (cf., e.g., \citep{lejaeghere_reproducibility_2016}).
However, if we simply apply periodic boundary conditions to $\AtomsNew$, the neighborhoods of atoms at the periodic domain boundaries would differ from those in the large-scale problem.
This is in particular severe for ``higher-than-zero-dimensional defects'', e.g., line (dislocations), surface (grain boundaries) or volume (cracks, voids) defects. The problem is thus  similar to hybrid quantum/classical mechanics (QM/MM) methods, where a small, but arbitrarily-shaped, cluster of atoms, treated with an ab initio model, is embedded in a much larger domain computed with interatomic potentials. To consider such clusters of atoms using plane-wave DFT, \citet{woodward_flexible_2002} developed two methods in the context of screw dislocations:
\begin{itemize}
 \item[(i)]
 In the first method, a vacuum region was introduced to separate the QM cluster from its periodic images. To prevent spurious free surface effects on cluster atoms, an additional buffer region replicating their true neighborhood was used in-between. 
 \item[(ii)]
 In the second method, the buffer region from method (i) was extended to the periodic domain boundaries to avoid the vacuum region. 
\end{itemize}
Method (ii) allows for a smaller cell size in comparison with method (i)---at the expense of a larger buffer region. However, neither method seems to be preferable over the other in terms of accuracy and computational cost since both have been equally successfully employed in many QM/MM studies (e.g., \citep{woodward_first-principles_2005,woodward_prediction_2008,bernstein_hybrid_2009,fellinger_geometries_2018}). In principle, they therefore also apply to our algorithm, however, none of them allows one to compute the quantum mechanical energy (of the cluster) which is a global quantity corresponding to the \emph{entire} periodic cell.\footnote{On the other hand, this may not be a problem if one is interested in purely ``force-based'' studies, e.g., computing defect geometries or the Peierls stress of a dislocation etc.}

\subsection{Symmetrization of extrapolative configurations}

Fortunately, in the case of MLIPs, we do not stringently require exact total energies---it suffices that the MLIP is able to \textbf{interpolate} on the large-scale configurational space. Nevertheless, it is still necessary to prevent free surfaces and/or other types of spurious boundary effects to avoid training on artificial neighborhoods irrelevant to the large-scale problem. We therefore present an improved version of method (ii) which symmetrizes the atomic displacements at the periodic domain boundaries to ensure that the neighborhood of the atoms outside the dislocation core is close to the one they would ``feel'' in the large-scale problem.

\begin{figure}[hbt]
 \centering
 \includegraphics[width=0.97\textwidth]{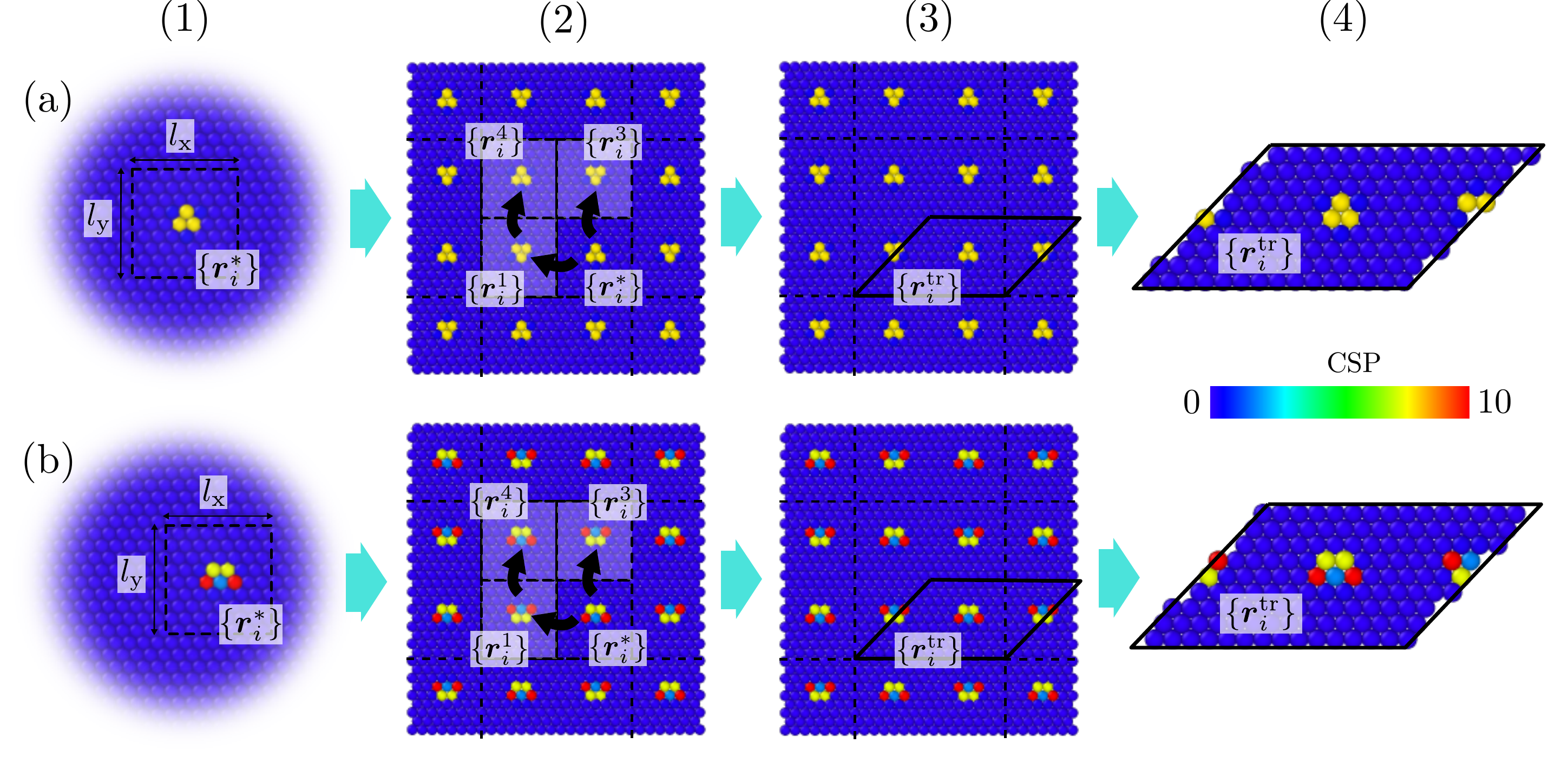}
 \caption{Schematic illustration of the individual steps (1)--(4) for constructing periodic training configurations for a screw dislocation (a) in the initial easy-core position and (b) shifted along the glide plane by a few lattice spacings with respect to (a). (1) Infinite configuration containing a single dislocation. (2) Periodic configuration after applying the displacement \eqref{eq:displ_mirr} to the reference lattice. (3) \& (4) Extraction of the dipole-like training configuration. Coloring of the atoms is due to the centrosymmetry parameter \citep[CSP,][]{kelchner_dislocation_1998}}
 \label{fig:construct_dft_cells}
\end{figure}

The idea is to extract the displacements $\tilde{\bdispl}_i(\Atom_{0,i}) = \Atom_i - \Atom_{0,i}$, where $\Atom_{0,i}$ is the ideal lattice site of the $i$-th atom, around the dislocation core and apply it to a periodic configuration, small enough to be computable by DFT but without introducing atomic neighborhoods different from those in the large-scale configuration.
To that end, we proceed as illustrated in Figure \ref{fig:construct_dft_cells} (a)-(1) and define the rectangular region of size $l_\rmx \times l_\rmy$, containing the dislocation core atoms $\AtomsNew$ which we check on extrapolation. Next, we define three replicas of this rectangular region and translate them according to Figure \ref{fig:construct_dft_cells} (a)-(2) by $l_\rmx$ with respect to the $\rmx$-axis and/or $l_\rmy$ with respect to the $\rmy$-axis.

Assuming that the dislocation line is parallel to the $\rmz$-axis, we then define the \emph{mirrored} displacement field
\begin{equation}\label{eq:displ_mirr}
 \bdispl(\bmx) =
 \left\{
 \begin{aligned}
  \;&\bdisplDD(x,y,z)                 &      &\bmx\in\{ \Atom_{0,i}^\ast \}, \\
    &\bdisplDD(-x-l_\rmx,y,z)         &      &\bmx\in\{ \Atom_{0,i}^1 \}, \\
    &\bdisplDD(x,-y-l_\rmy,z)         &      &\bmx\in\{ \Atom_{0,i}^2 \}, \\
    &\bdisplDD(-x-l_\rmx,-y-l_\rmy,z) &\qquad&\bmx\in\{ \Atom_{0,i}^3 \},
 \end{aligned}
 \right.
\end{equation}
where $\{ \Atom_{0,i}^1 \}$--$\{ \Atom_{0,i}^3 \}$ are the ideal lattice sites in the translated regions. Imposing this displacement gives the configurations $\{ \Atom_i^1 \}$--$\{ \Atom_i^3 \}$ shown in Figure \ref{fig:construct_dft_cells} (a)-(2). The union of $\AtomsNew$ and $\{ \Atom_i^1 \}$--$\{ \Atom_i^3 \}$ is clearly periodic and from the figure it can be seen that this procedure creates minimal spurious effects in the sense that the atomic neighborhoods are not ``too far'' from the large-scale problem, indicated by the centrosymmetry parameter which is less than 0.04 in the vicinity of the periodic domain boundaries---which is more than two orders of magnitude smaller than for the core atoms.
This configuration appears similar to the well-known quadrupolar cell (cf., e.g., \citep{bigger_atomic_1992,samolyuk_influence_2013}), yet it is emphasized that our configuration fully includes the elastic far-field contribution due to external boundary conditions intrinsic to the large-scale problem.

We may further reduce this cell to the dipole-like configuration framed by the continuous line in Figure \ref{fig:construct_dft_cells} (a)-(3) along similar lines as for the quadrupolar cells if the displacement is point symmetric with respect to the dislocation positions. To see why this choice is indeed justified we refer to our analysis in \ref{sec:dipole_config_analysis}.

It should be noted that this construction applies independently of the position of the dislocation core as shown in Figure \ref{fig:construct_dft_cells} (b). That is, if the dislocation starts moving, e.g., due to some far-field applied stress, we detect the dislocation core position and proceed in the same way as above by shifting the rectangular region accordingly.

Even though the proposed symmetrization technique ensures that no artificial neighborhoods occur in the training configuration, we would like to point out that there is potential freedom in modifying the positions of the boundary atoms: if, e.g., the symmetry of the displacement field breaks due to additional long-range effects, one may introduce an intermediate step which minimizes the centrosymmetry parameter, or, alternatively, the extrapolation grade, of atoms residing in some boundary layer.

\begin{rem2*}
 We find it important to emphasize that, despite the fact that the dislocation lines are extremely (some may say ``unphysically'') close to each other in the training set, the fitted model is still expected to (and does in fact, as we will see in next section) give accurate predictions.
 The reason for such training sets being representative of configurations with normal, ``physical'' distance between dislocations rests on the main assumption behind the success of machine-learning potentials: that the perturbations to electronic degrees of freedom decrease fast and therefore the electronic interaction can be well-described by the relative positions of atoms in a small neighborhood (like 5\,\AA).
 Hence, if the atomic neighborhoods in the training set are representative of those in a large-scale simulation, then the potential is expected to give accurate predictions.
 The difference in the long-range elastic energy between training and actual configurations will be resolved by the atomistic deformation which is explicitly represented in the problem of interest.
 
	 	In particular, total energies and forces of the small periodic training configurations are not used to compute any physical quantities (defect formation energies, energy barriers etc.).
	 	On the contrary, they are computed in the large-scale simulation in which we are free to choose any boundary condition (e.g., flexible boundary conditions \citep{sinclair_flexible_1978,woodward_flexible_2002,ehrlacher_analysis_2016,hodapp_lattice_2019}).
	 	Hence, there is no need for the application of correction methods (e.g., \citep{varvenne_elastic_2017,ma_calanie_2020}) employed when operating solely on small, fully periodic DFT cells due to the interaction of defects with their periodic images \textbf{through elastic fields}.
 	This is another advantage of our method.
\end{rem2*}

\section{Computational results}
\label{sec:cyl}

We now test our active learning algorithm for $1/2\,\langle 111 \rangle$ screw dislocation motion in bcc tungsten in order to predict the core structure, the Peierls barrier and the Peierls stress. Our large-scale problem is a cylindrical configuration of atoms and we choose the region where we check extrapolation of the MTP as the collection of atoms in the rectangular region with dimensions $l_\rmx = 5 \sqrt{3}\latConst/\sqrt{2}$ and $l_\rmx = 9 \latConst/\sqrt{2}$, where $\latConst$ denotes the lattice constant, centered at the dislocation core (cf. Figure \ref{fig:construct_dft_cells} (1)). From this extrapolation region we construct the training set, as proposed in the previous section, leading to a total number of 135 atoms in the periodic configurations (i.e., in $\{ \Atom_i^{\rm tr} \}$, cf. Figure \ref{fig:construct_dft_cells} (4)). As an ab initio model, we employ DFT by means of the Vienna ab initio simulation package \citep[VASP,][]{kresse_efficient_1996} which uses plane-wave basis sets and the projector-augmented wave (PAW) pseudopotential method \citep{blochl_projector_1994,kresse_ultrasoft_1999}. For all DFT calculations we have used the simulation parameters presented in \ref{sec:VASP}.

For our large-scale atomistic simulations we use nonlinear MTPs as described in \ref{sec:mtp_nln}. Since we focus on applications involving geometry optimization and barrier calculations we fit the MTPs to energies and forces according to the procedure described in \ref{sec:mtp_train}. The parameter $N$, that is, the number of iterations we perform before we retrain the MTPs (see the algorithm in Section \ref{sec:algo}), is of the order of 30 for all our simulations.

The size of the cylindrical configuration is $\approx\,$5700 atoms for the core relaxation and the Peierls stress calculation in Section \ref{sec:cyl.relax} and \ref{sec:cyl.peierls_stress}, respectively. For the barrier calculations in Section \ref{sec:cyl.neb} the size is $\approx\,$1400 atoms. We compare the results obtained by using the cylindrical configuration to the results obtained by using the classical 135-atom dipole configuration (e.g., \citep{ventelon_ab_2013}). In the following we therefore refer to them as the cylinder and the periodic problem, respectively.

For validation purposes, we have also trained the MTP with respect to two empirical interatomic potentials based on the embedded atom method (EAM), namely the EAM3 and EAM4 potentials developed in \citep{marinica_interatomic_2013}.%
\footnote{Versions of both potentials were obtained from the Knowledgebase of Interatomic Models \citep[KIM,][]{elliott_knowledgebase_2011} repository}
For clarity, we therefore refer to the MTP trained with respect to DFT as MTP-DFT and the MTPs trained with respect to the EAM potentials as MTP-EAM3 and MTP-EAM4, respectively. For compactness, we only present a detailed discussion of the results for the large-scale cylinder problem using the MTP-DFT, which is our main contribution, and refer to the results of the validation (such as training errors etc.) for the MTP-EAM3 and the MTP-EAM4, and also the MTP-DFT for the periodic problems, to \ref{sec:train_err_mtp-eamX}.

In our test problems we consider solely structural relaxation using the fast inertial relaxation engine \citep[FIRE,][]{bitzek_structural_2006} as implemented in the atomic simulation environment \citep[ASE,][]{hjorth_larsen_atomic_2017} library. The active learning algorithm is terminated when the maximum force on an atom is less than 0.001\,eV/\AA; except for the barrier calculations for which we use 0.01\,eV/\AA.

\subsection{Core relaxation}
\label{sec:cyl.relax}

For the dislocation core relaxation we set the bounds $\gamma_{\rm min}$ and $\gamma_{\rm max}$, within which we consider configurations as potential candidates for the training set, to 1.1 and 10, respectively. This choice is close to the optimum (cf. \citep{podryabinkin_active_2017}) and should therefore yield the best possible accuracy that one could achieve with an MTP in practice. Using this parameter set, the active learning algorithm converged after 63 iterations with a training set size of 40 configurations.
The training errors reported in Table \ref{tab:screw_cyl_training_err} are in the range of the numerical approximation of DFT codes themselves. Therefore, the MTP essentially reproduces the training data without introducing further errors. 

\begin{table}[t]
 \renewcommand{\arraystretch}{1.2}
 \centering
 \begin{tabular}{c||c|c|c}
  & \multicolumn{3}{c}{Error in:} \\
  & Energy [eV] & Energy per atom [eV] & Forces [eV/\AA] \\
  \hline\hline
  Maximal absolute difference & 8.17$\cdot$10$^{-4}$ & 6.052$\cdot$10$^{-6}$ & 0.0284 \\ \hline
  Average absolute difference & 4.069$\cdot$10$^{-4}$ & 3.015$\cdot$10$^{-6}$ & 0.0112 \\ \hline
  RMS                         & 4.739$\cdot$10$^{-4}$ & 3.51$\cdot$10$^{-6}$ & 0.0121 \\ \hline\hline
  Max(ForceDiff) / Max(Force) & & & 0.0284 \\ \hline
  RMS(ForceDiff) / RMS(Force) & & & 0.0494
 \end{tabular}
 \caption{Training errors corresponding to the core relaxation for the MTP-DFT. The errors are of the same order than the numerical precision of our (already conservative) DFT calculation (cf. \ref{sec:VASP}) and, consequently, the MTP-DFT is able to reproduce the DFT model practically without error.}
 \label{tab:screw_cyl_training_err}
\end{table}

\begin{figure}[t]
 \centering
 \includegraphics[width=0.28\textwidth]{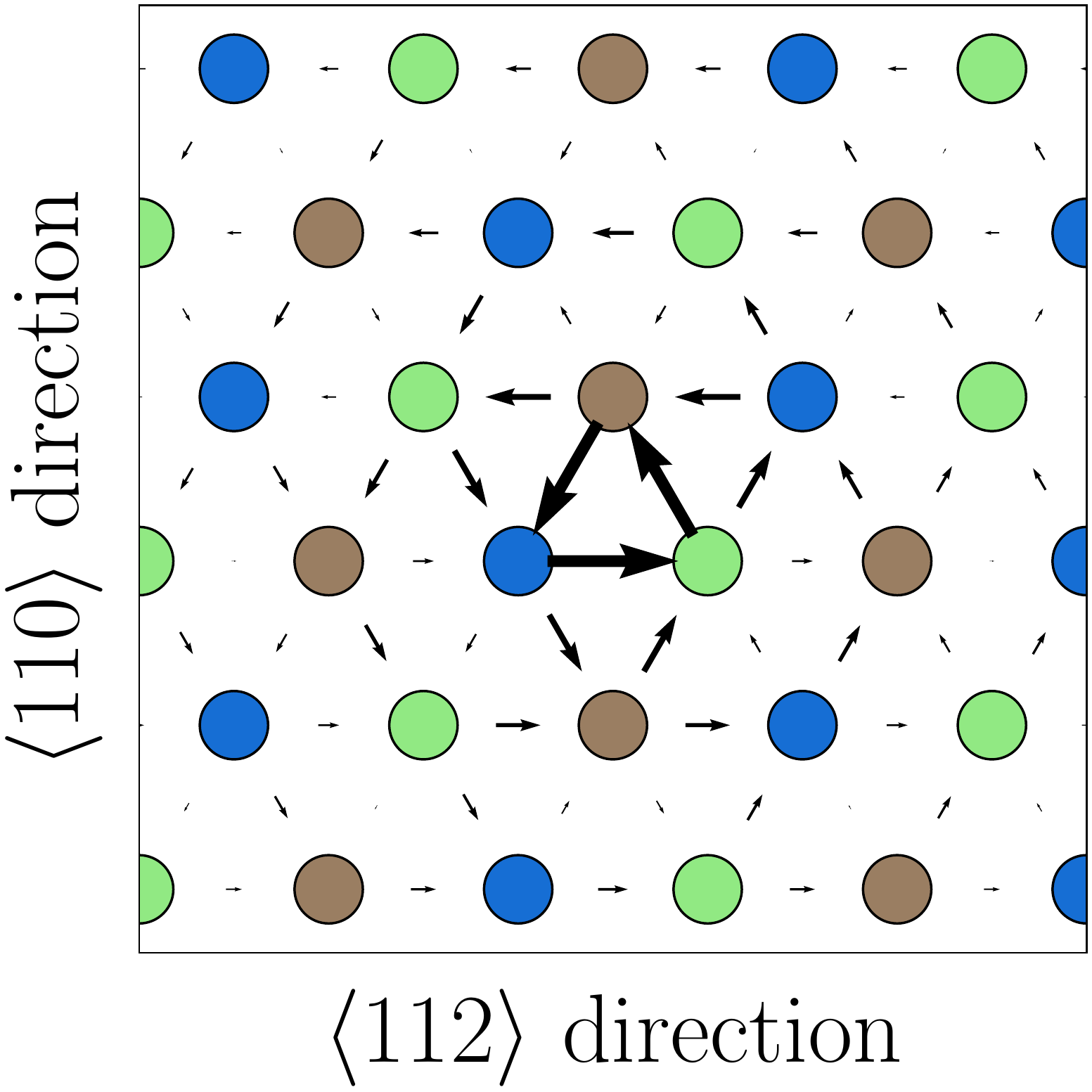}
 \caption{Differential displacement map around the dislocation core after relaxation for the MTP-DFT which displays the compact structure in agreement with DFT}
 \label{fig:screw_cyl_ddm}
\end{figure}

The differential displacement map for the relaxed core structure computed with the MTP is shown in Figure \ref{fig:screw_cyl_ddm}. It can immediately be seen that our algorithm predicts the compact core structure from previous established DFT studies (cf., e.g., \citep{samolyuk_influence_2013,ventelon_ab_2013}). To quantify this result, we show the energy difference between the initial linear elastic and the final relaxed configuration in Table \ref{tab:screw_cyl_Ediff}. This is of course not feasible to compute with a reference DFT calculation so that we compare our result to the 135-atom periodic configuration. The energy difference is $\approx\,$2$\times$ larger for the periodic problem, which is expected since there are two dislocations in this system. In addition, we have computed the energy difference for the screw-cylinder configuration using the EAM4 potential as a reference model---where we \emph{can} compute the exact solution. Here, the error of the MTP-EAM4 is $\approx\,$2\% which confirms the excellent agreement between both models and provides further evidence that our algorithm reliably extracts the core energy; note that an error of 1--2\,\% also occurs for the periodic problem (compare the EAM4 and MTP-EAM4 values in Table \ref{tab:screw_cyl_Ediff}).

\begin{table}[t!]
 \renewcommand{\arraystretch}{1.2}
 \centering
 \begin{tabular}{c||c|c|c|c}
  & DFT & MTP-DFT & EAM4 & MTP-EAM4 \\
  \hline\hline
  periodic & $-0.298$ & $-0.298$ & $-0.822$ & $-0.813$ \\ \hline
  cylinder &  n/a  & $-0.166$ & $-0.415$ & $-0.407$
 \end{tabular}
 \caption{Energy differences (in eV per Burgers vector) between the initial and final (relaxed) configurations. The MTP-DFT \emph{exactly} reproduces the DFT value for the periodic problem while the small discrepancy between the MTP-EAM4 and the EAM4 is likely due to the nonsmooth potential energy surface (cf. Figure \ref{fig:potential_functions_eam4}. Since this discrepancy is the same for both the periodic and the cylinder problem, and, moreover, the fitting errors are in the range of DFT precision (cf. Table \ref{tab:screw_cyl_training_err}), the MTP-DFT presumably also \emph{exactly} reproduces the (infeasible) cylinder DFT value}
 \label{tab:screw_cyl_Ediff}
\end{table}

\begin{figure}[hbt]
 \centering
 \includegraphics[width=0.9\textwidth]{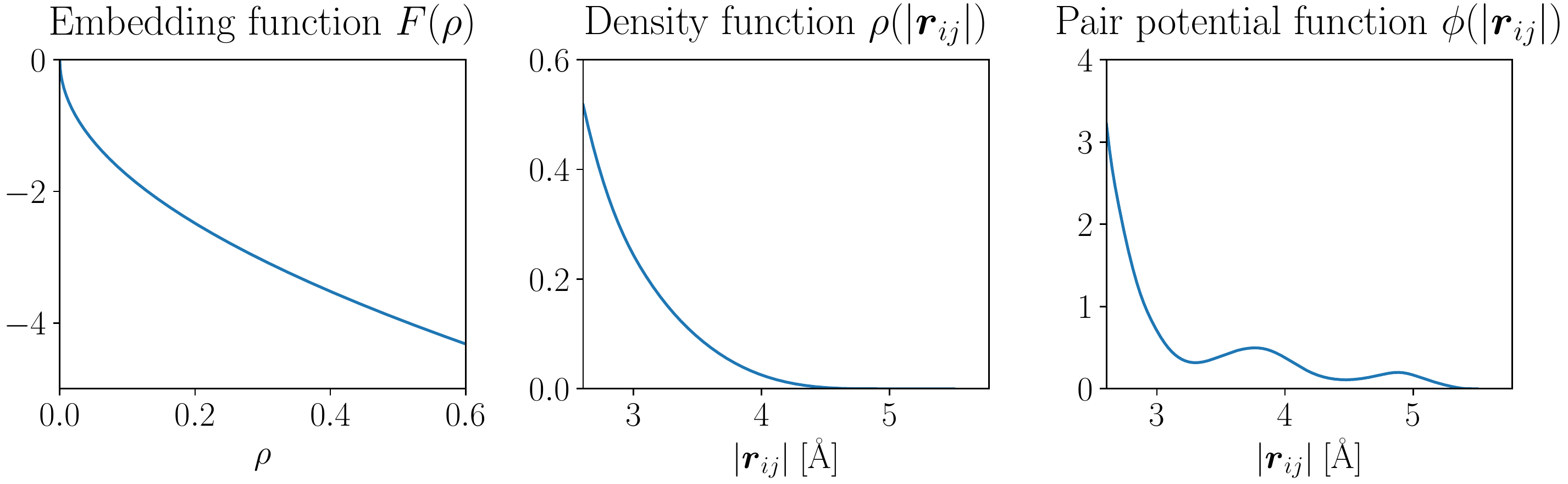}
 \caption{Potential functions of the EAM4 per-atom energy functional $\Esite(\Neigh) = \sum_{\Atom_{ij}} \mphi(\abs{\Atom_{ij}}) + F(\sum_{\Atom_{ij}} \mrho(\abs{\Atom_{ij}}))$. The ranges of the function arguments are adapted to the atomic neighborhoods occurring during the simulation}
 \label{fig:potential_functions_eam4}
\end{figure}

It is interesting to note that it was ``easier'' for the MTP to reproduce DFT than EAM4---the difference between MTP and DFT is less than 1\,meV per Burgers vector, while the difference between MTP and EAM4 is of the order of 10\,meV per Burgers vector.
Speculatively, we attribute it to the fact that the MTP has a flexible functional form which makes it easy to reproduce the smooth underlying DFT energy (at least, when the configurational space is not too large), whereas fitting an EAM potential introduces an oscillatory behavior commensurate with atomic shell radii in the pair potential function as shown in Figure \ref{fig:potential_functions_eam4}.
Such a nonsmooth behavior appears to be more difficult to fit with a smooth MTP radial basis than fitting the DFT energy.

\subsection{Peierls barrier}
\label{sec:cyl.neb}

Using the relaxed configuration from the previous section, we now compute the Peierls barrier for screw dislocation motion under zero stress by calculating the minimum energy path using the nudged elastic band \citep[NEB,][]{henkelman_improved_2000} method. We create the final image by translating the displacement field of the relaxed core structure by $\sqrt{2/3}\,\latConst$ along the $\langle 112 \rangle$ direction, i.e., to the next easy-core position on the $\{110\}$ plane. To create the initial guess for the intermediate images, we use the same displacements according to their reaction coordinate.

For this problem we have chosen the extrapolation bounds $\gamma_{\rm min}=2$ and $\gamma_{\rm max}=10$. Using an NEB with 5 images, the active learning algorithm converged after 38 iterations with a training set size of 22 configurations. Therefore, our algorithm \emph{only} required a full DFT calculation on the 135-atom symmetrized core configuration (cf. Figure \ref{fig:construct_dft_cells} (4)) roughly every 9th iteration.
The training errors in Table \ref{tab:screw_cyl_neb_training_err} are slightly worse than for the core relaxation, which is expected since the training set is now more diverse due to the intermediate core configurations. However, a maximum error in the total energy of 5.6\,meV (per configuration) is still a very small quantity since the Peierls barrier is more than an order of magnitude higher (cf. Table \ref{tab:screw_neb_energies}).

\begin{table}[b!]
 \renewcommand{\arraystretch}{1.2}
 \centering
 \begin{tabular}{c||c|c|c}
  & \multicolumn{3}{c}{Error in:} \\
  & Energy [eV] & Energy per atom [eV] & Forces [eV/\AA] \\
  \hline\hline
  Maximal absolute difference & 0.0056 & 4.152$\cdot$10$^{-5}$ & 0.0756 \\ \hline
  Average absolute difference & 0.00233 & 1.728$\cdot$10$^{-5}$ & 0.0176 \\ \hline
  RMS                         & 0.00274 & 2.029$\cdot$10$^{-5}$ & 0.0199 \\ \hline\hline
  Max(ForceDiff) / Max(Force) & & & 0.0786 \\ \hline
  RMS(ForceDiff) / RMS(Force) & & & 0.0647
 \end{tabular}
 \caption{Training errors corresponding to the NEB calculation for the MTP-DFT. The maximum error in the energy is of the order of a few meV which confirms that the training set consisting of the symmetrized configurations are suitable for representing intermediate core structures (cf. Figure \ref{fig:construct_dft_cells} (b))}
 \label{tab:screw_cyl_neb_training_err}
\end{table}

\begin{table}[b!]
 \renewcommand{\arraystretch}{1.2}
 \centering
 \begin{tabular}{c||c|c|c|c}
  & DFT \citep{grigorev_hybrid_2020} & MTP-DFT & EAM4 & MTP-EAM4 \\
  \hline\hline
  periodic & 0.0921 & 0.0905 & 0.0589 & 0.0570 \\ \hline
  cylinder & n/a & 0.0880 & 0.0627 & 0.0629
 \end{tabular}
 \caption{Peierls energies in eV per Burgers vector corresponding to the energy curves in Figure \ref{fig:screw_neb_energy_curves_&_ddm} (a). The small difference of $\approx$\,1\,meV per Burgers vector and less between the available MTP-DFT/MTP-EAM4 and DFT/EAM4 values coincides with the training errors (see \ref{sec:training_errors_neb}) and is thus also expected between the MTP-DFT and DFT for the cylinder problem (cf. Table \ref{tab:screw_cyl_neb_training_err})}
 \label{tab:screw_neb_energies}
\end{table}

\begin{figure}[t]
 \begin{minipage}[t]{0.65\textwidth}
  \centering
  (a)\\[0.5em]
  \includegraphics[width=0.9\textwidth]{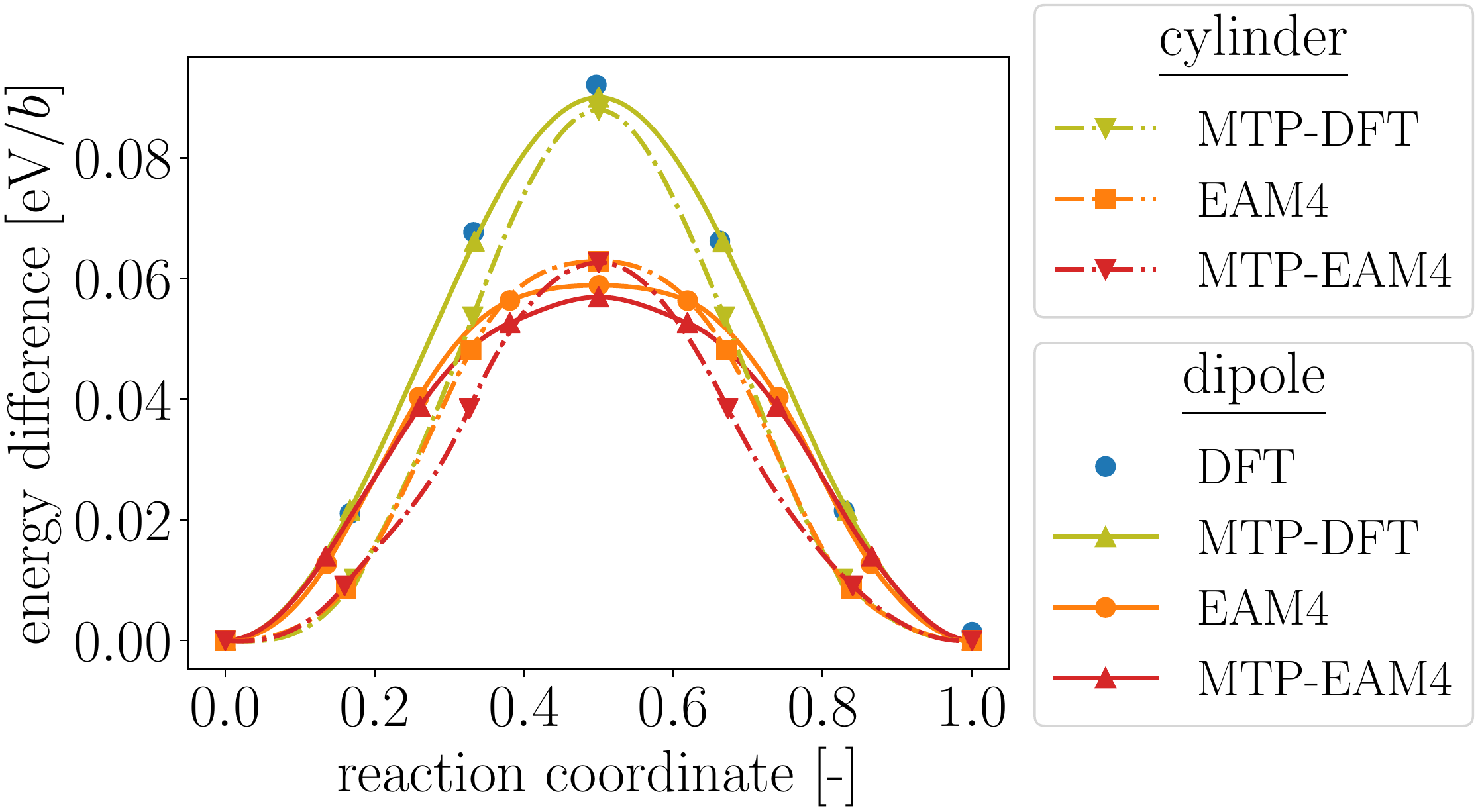}
 \end{minipage}\hfill
 \begin{minipage}[t]{0.35\textwidth}
  \centering
  (b)\\[1.8em]
  \includegraphics[width=0.8\textwidth]{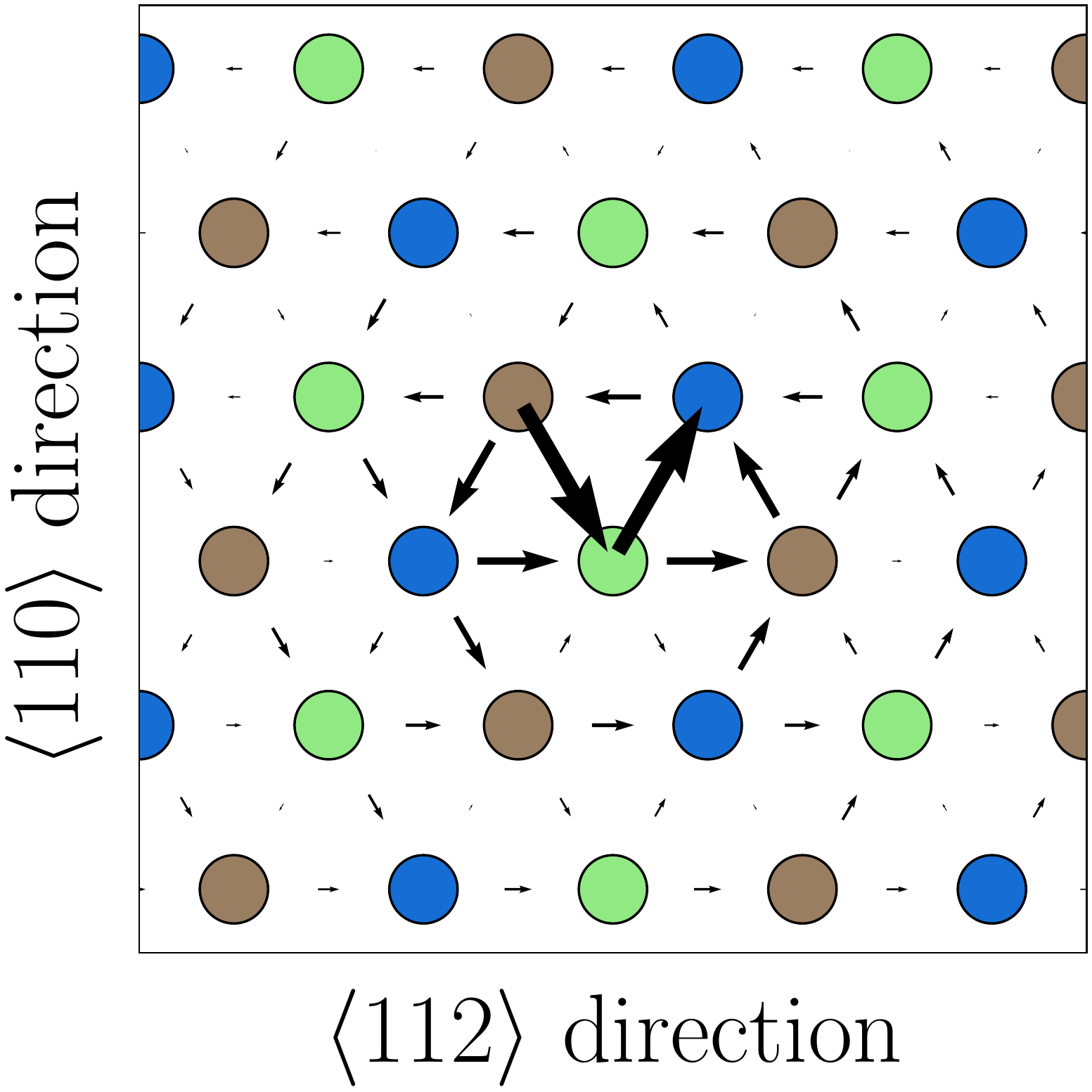}
 \end{minipage}
 \caption{(a) Energy curves for screw dislocation motion. The deviations of all MTP-X curves from the exact values are tiny (of the order of 1\,meV per Burgers vector!) and, therefore, DFT as well as EAM4 are very-well reproduced by the MTP. Note that the small differences between the curves for the cylinder and periodic problems are expected since the elastic interaction between the dislocation cores largely cancels out due to the quadrupole arrangement in the periodic case (cf. \citep{grigorev_hybrid_2020}). (b) Differential displacement map around the dislocation core of the central image for the MTP-DFT which shows the favorable intermediate configuration also predicted by DFT}
 \label{fig:screw_neb_energy_curves_&_ddm}
\end{figure}

The corresponding energy curve is shown in Figure \ref{fig:screw_neb_energy_curves_&_ddm} (a). Again, since a reference DFT calculation is unfeasible, we have compared our result to the one obtained from a periodic calculation using solely DFT. According to the convergence study by \citet{grigorev_hybrid_2020}, the difference between a periodic configuration of 135 atoms and an effectively infinite configuration should be in the range of only a few percent. From Table \ref{tab:screw_neb_energies}, it follows that the difference is $\approx\,$4\% and thus in agreement with their analysis. For testing purposes, we have also carried out calculations with the EAM4 potential as our reference model, as in the previous Section. Here, the energy difference between the periodic and the cylinder configuration is $\approx\,$6\%. Moreover, the Peierls energies for the EAM4 and the MTP-EAM4 coincide almost perfectly which further validates our training methodology. In addition, the differential displacement map for the relaxed central image is shown in Figure \ref{fig:screw_neb_energy_curves_&_ddm} (b) which demonstrates that the MTP-DFT also correctly predicts the same intermediate configuration observed in our and recent DFT studies.

\subsection{Peierls stress}
\label{sec:cyl.peierls_stress}

Finally, we test our algorithm for the screw-cylinder problem subject to a far-field applied stress in order to predict the Peierls stress, that is, the stress under which the dislocation starts moving. To this end, we apply a deformation corresponding to the shear stress $\stressShear_{\rm yz}$ to all atoms and subsequently relax the system. During relaxation, we detect the position of the dislocation core and update the extrapolation region accordingly (cf. Figure \ref{fig:construct_dft_cells} (b)). For this problem, we have used $\gamma_{\rm min}=2$ and $\gamma_{\rm max}=10$ as for the barrier calculations.

We have applied an initial stress of $\approx$\,1.74\,GPa and increased it in 0.1\,GPa-steps upon convergence. The dislocation then moved to the next easy-core position when the stress was 2.34\,GPa. Hence, we define the obtained Peierls stress as 2.29$\,\pm\,$0.05\,GPa. This value is in the same range as previous DFT studies as shown in Table \ref{tab:peierls_stress}. Nevertheless, we would like to remind the reader that these previous studies use \emph{indirect} methods; e.g., the method in \citep{dezerald_plastic_2016} is based on fitting a Kocks-type function using the Peierls barriers from several periodic calculations subject to different applied stresses as fitting parameters. They are therefore computationally significantly more expensive, require prior knowledge of the slip plane and are also prone to errors due to the analytic fit.

\begin{table}[b]
 \renewcommand{\arraystretch}{1.2}
 \centering
 \begin{tabular}{c|c|c|c|c|c}
  DFT \citep{romaner_effect_2010} & DFT \citep{samolyuk_influence_2013} & DFT \citep{dezerald_plastic_2016} & MTP-DFT & EAM3 & MTP-EAM3 \\
  \hline\hline
  2.5--2.8 & 1.71 & 2 & 2.29$\,\pm\,$0.05 & 0.97$\,\pm\,$0.005 & 0.904$\,\pm\,$0.005
 \end{tabular}
 \caption{Peierls stress (in GPa) for W predicted by DFT and the interatomic potentials. The MTP-DFT value lies well within the range of all considered previous DFT studies. It is remarked that, given the error in the Peierls stress for the MTP-EAM3 (which has similar training errors compared to the MTP-DFT) with respect to the EAM3 potential which is at most 8\,\%, the anticipated exact DFT value (for our parameter set) lies between 2 and 2.5\,GPa}
 \label{tab:peierls_stress}
\end{table}

In order to judge on the accuracy of our method, we compute the Peierls stress with the EAM3 potential as the reference model which allows us to perform an exact large-scale calculation.\footnote{Here, we have used the EAM3 potential since the EAM4 potential does not predict the $\{ 110 \}$ glide plane} Following the same procedure as described above, except that we now increment the applied stress by 0.01\,GPa, the predicted Peierls stress for the MTP-EAM3 agrees well with our large-scale reference calculation showing an error of only 6--8\% (cf. Table \ref{tab:peierls_stress}). Moreover, this corresponds to the relative fitting errors for the MTP-EAM3 shown in Table \ref{tab:screw_cyl_strain_training_err_eam3}. Given that the training errors for the MTP-EAM3 and the MTP-DFT (Table \ref{tab:screw_cyl_strain_training_err}) are in the same range, it is likely that the DFT Peierls stress can also be predicted with the MTP-DFT up to an error of 6--8\%. Should we need a higher accuracy, we may simply increase the number of basis functions of the MTP but note as well that errors up to 10\% are absolutely tolerable when the Peierls stress is used to calibrate a higher-scale model, such as discrete dislocation dynamics.

\begin{table}[t!]
 \renewcommand{\arraystretch}{1.2}
 \centering
 \begin{tabular}{c||c|c|c}
  & \multicolumn{3}{c}{Error in:} \\
  & Energy [eV] & Energy per atom [eV] & Forces [eV/\AA] \\
  \hline\hline
  Maximal absolute difference & 0.0233 & 1.723$\cdot$10$^{-4}$ & 0.133 \\ \hline
  Average absolute difference & 0.00806 & 5.969$\cdot$10$^{-5}$ & 0.0177 \\ \hline
  RMS                         & 0.00955 & 7.073$\cdot$10$^{-5}$ & 0.0198 \\ \hline\hline
  Max(ForceDiff) / Max(Force) & & & 0.123 \\ \hline
  RMS(ForceDiff) / RMS(Force) & & & 0.0644
 \end{tabular}
 \caption{Training errors corresponding to the Peierls stress calculation for the MTP-DFT. The errors are not significantly higher than for the barrier calculations (cf. Table \ref{tab:screw_cyl_neb_training_err}) although the training data now consists of a more diverse set of configurations subject to various applied stresses within the range 1.74--2.34\,GPa. It is therefore expected that the MTP-DFT also predicts the core energies for strained configurations without any significant loss of accuracy when compared to the results in Section \ref{sec:cyl.relax} and \ref{sec:cyl.neb}}
 \label{tab:screw_cyl_strain_training_err}
\end{table}

\begin{figure}[t]
 \centering
 \begin{minipage}[t]{0.45\textwidth}
  \centering
  (a)\\[0.5em]
  \includegraphics[width=0.95\textwidth]{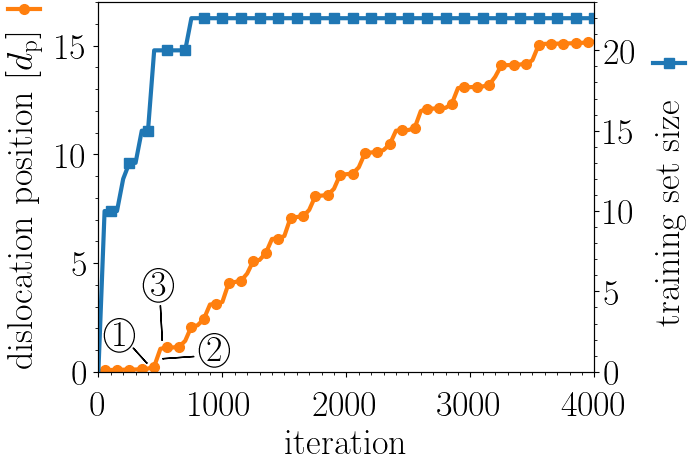}
 \end{minipage}\hfill
 \begin{minipage}[t]{0.55\textwidth}
  \centering
  (b)\\[1.3em]
  \includegraphics[width=0.9\textwidth]{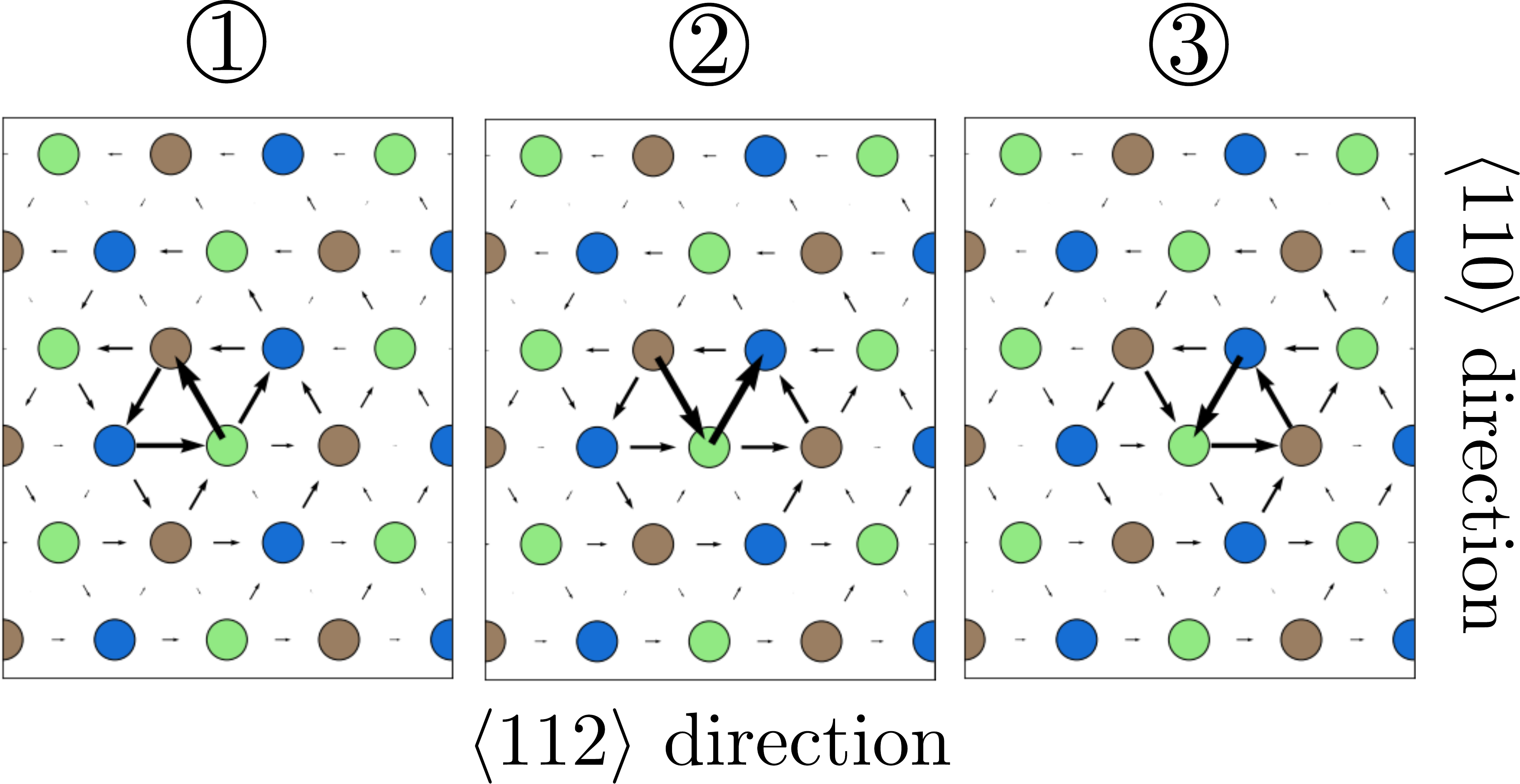}
 \end{minipage}
 \caption{(a) Dislocation position in $d_\rmp = \sqrt{2/3}\,a_0$, where $d_\rmp$ is the distance between two easy-core positions, and training set size as functions of the iteration for the Peierls stress calculation from Section \ref{sec:cyl.peierls_stress} using the MTP-DFT. The figure demonstrates that the active learning algorithm works properly by showing that no further configurations are added to the training set after the dislocation has crossed the first Peierls valley after $\approx$\,700\;iterations (and henceforth only encounters ``known'' neighborhoods). (b) Visualization of the dislocation core structure using a differential displacement map at selected iterations from (a), i.e., \raisebox{.5pt}{\textcircled{\raisebox{-0.1em}{1}}} before, \raisebox{.5pt}{\textcircled{\raisebox{-0.1em}{2}}} during and \raisebox{.5pt}{\textcircled{\raisebox{-0.1em}{3}}} after crossing the first Peierls barrier; thereby, the dislocation core structure correctly changes from easy to intermediate, and back again to easy, without admitting any other configuration (which also explains the small training set size of 22 configurations shown in (a))}
 \label{fig:screw_cyl_strain_history_&_ddm}
\end{figure}

To validate the correct functioning and demonstrate the capabilities of the active learning algorithm, we show the evolution of the dislocation position (with respect to the origin) and the training set size as a function of the iteration in Figure \ref{fig:screw_cyl_strain_history_&_ddm} (a). Having crossed the first Peierls barrier after $\approx 550$ iterations, the training set contained 20 configurations, increased to 22 while crossing the second Peierls barrier, and then continued gliding without triggering new DFT calculations. This behavior is expected since the dislocation moves by alternating from \raisebox{.5pt}{\textcircled{\raisebox{-0.1em}{1}}} easy-core $\rightarrow$ \raisebox{.5pt}{\textcircled{\raisebox{-0.1em}{2}}} intermediate-core $\rightarrow$ \raisebox{.5pt}{\textcircled{\raisebox{-0.1em}{3}}} easy-core $\rightarrow$ ... etc. as shown in Figure \ref{fig:screw_cyl_strain_history_&_ddm} (b) and does therefore not encounter any significantly new, i.e., extrapolative, configurations.

\section{Concluding remarks}
\label{sec:conclusions}

\subsection{Summary and discussion}

We have developed an active learning algorithm for large-scale atomistic simulations using moment tensor potentials (MTPs), a class of machine-learning interatomic potentials (MLIPs). Our algorithm allows for a novel way of performing large-scale simulations with DFT accuracy fully automatically by letting active learning find the right training data \emph{during} the simulations---without the need for any prior (passive) training. We have applied the algorithm to model screw dislocation motion in bcc tungsten and shown that the MTP is able to reproduce known mechanically relevant properties from the literature, such as core structure, transition state barriers and Peierls stress, alongside with a significant reduction in the necessary amount of DFT calculations compared to standard methods.

More specifically, we have shown that active learning is able to reliably detect a minimum set of sufficiently distinct training configurations. This is in particular demonstrated in Section \ref{sec:cyl.peierls_stress}, where the dislocation moves through the effectively infinite crystal: in the beginning, active learning requests new training data during the change of the core structure but stops when the dislocation repetitively encounters the same configurations after crossing the first Peierls valley. Moreover, using our symmetrization strategy to convert extrapolative regions into periodic configurations, the MTP gave accurate predictions for forces as well as energy differences.

Our algorithm can in principle be applied by a user in the same way as a standard large-scale atomistic simulation. Therefore, no additional expert knowledge is required in order to extract physical quantities, such as the Peierls stress, from indirect DFT methods (cf. Section \ref{sec:cyl.peierls_stress}). This opens new prospects for using MLIPs in general as a robust tool for DFT-accurate simulations of extended defects.

\subsection{Outlook}

Our example in Section \ref{sec:cyl.peierls_stress} further shows that we can include far-field applied stresses which influence the core structure of the dislocation and its the elastic field. We therefore expect that our algorithm is directly applicable to related problems involving interactions between dislocations and different types of defects, e.g., vacancies or other dislocations.

While we have assumed that nonlinearities are confined to the small-sized extrapolation region(s), we have demonstrated that active learning significantly reduces the amount of actual DFT calculations. If the size of the defect core increases to several hundreds of atoms, e.g., in the case of dissociated dislocations in face-centered-cubic lattices, the saved computing capacities may thus be invested in a few larger calculations in order to compute energies; combined with many small cluster calculations in order to compute forces. 

\begin{figure}[hbt]
 \centering
 \includegraphics[width=0.8\textwidth]{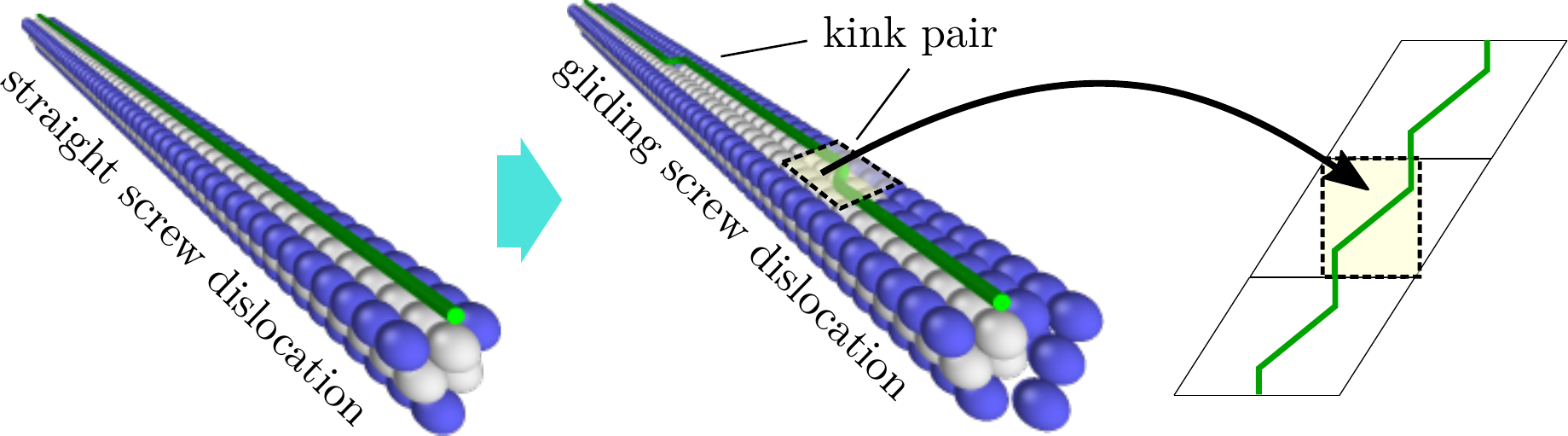}
 \caption{Schematic illustration of a possible method for constructing periodic training configurations for dislocation kinks}
 \label{fig:kink_nucleation}
\end{figure}

Further, we think that our methodology will also be useful for inherently three-dimensional problems, e.g., curved dislocations. For this class of problems, we currently envision two possible approaches. The first possibility is to train a MLIP on several selected straight dislocations with different character angles. The idea is then to switch off active learning and use the MLIP as a standard interatomic potential. The second possibility is to proceed as we do in the present work, i.e., we carve out a collection of atoms in the vicinity of the dislocation core and construct a periodic configuration. This procedure is illustrated in Figure \ref{fig:kink_nucleation} for a three-dimensional dislocation in a bcc lattice moving by means of the kink-pair nucleation mechanism (see, e.g., \citep{hirth_theory_1982}). Both directions will be explored in future work.

It is finally worthwhile to emphasize that our methodology does not exclude multi-component materials involving alloying elements and chemical impurities, e.g., hydrogen or helium, as most MLIPs already include the necessary descriptors (e.g., \citep{behler_generalized_2007,gubaev_accelerating_2019}). The discovery of new materials mainly comes through the search for novel alloys and we think that investigating corresponding mechanisms using MLIPs in combination with active learning will be one of the most auspicious potential applications since the configurational space will likely be too vast to be explored with only passively selected training sets.


\section*{Acknowledgements}

We thank Dr. Evgeny Podryabinkin for assistance during the implementation of the neighborhood selection criterion. Moreover, MH would like to thank Dr. Petr Grigorev for helpful discussions regarding the DFT calculations for screw dislocations.

Financial support from the Russian Science Foundation (grant number 18-13-00479) and the Fonds National Suisse (FNS), Switzerland, (project 191680) is highly acknowledged.


\begin{appendix}

\section{Moment tensor potentials}
\label{sec:mtp}

\subsection{Potential representation}
\label{sec:mtp_rep}

The basis functions from \eqref{eq:Eatom} for MTPs are given by the following form \citep{shapeev_moment_2016}
\begin{equation}
 B_\alpha(\Neigh) = \prod_{l=1}^k M_{\mu_l(\alpha),\nu_l(\alpha)}(\Neigh),
\end{equation}
where
\begin{equation}
 M_{\mu_l,\nu_l}(\Neigh) = \sum_{\Atom_{ij} \in \Neigh} f_{\mu_l}(\abs{\bmr_{ij}})
 \, \underbrace{\bmr_{ij} \otimes \cdots \otimes \bmr_{ij}}_{\nu_l \; \text{times}},
\end{equation}
are the moment tensor descriptors. Here, the $f_{\mu_l}$'s are radial basis functions which vanish for $\abs{\bmr_{ij}} > r_{\rm cut}$, where $r_{\rm cut}$ is a predefined cut-off radius. We characterize the degree of an MTP by its level. That is, an MTP of level $d$ implies that the $B_\alpha$'s comprise all basis functions which satisfy $(2\mu_1 + \nu_1) + (2\mu_2 + \nu_2) + ... \le d$.

For all numerical examples, conducted in this work, we have used an MTP of level $d=16$ with a cut-off radius $r_{\rm cut}=5$\,\AA.

\subsection{Nonlinear MTPs}
\label{sec:mtp_nln}

In the case of nonlinear MTPs, as we use them in the current work, the radial basis functions $f_{\mu_l}$ depend on an additional parameter vector $\uc_{\mu_l}$ as follows (cf. \citep{gubaev_accelerating_2019})
\begin{equation}
 f_{\mu_l}(\abs{\bmr_{ij}};\uc_{\mu_l}) = \sum_n c^n_{\mu_l} Q^n(\abs{\bmr_{ij}}), \qquad \text{with}
 \; Q^n(\abs{\bmr_{ij}}) = T_n(\abs{\bmr_{ij}})(r_{\rm cut} - \abs{\bmr_{ij}})^2,
\end{equation}
where $T_n$ is the $n$-th order Chebysev polynomial and the function $(r_{\rm cut} - \abs{\bmr_{ij}})^2$ ensures a smooth transition to 0 as $\abs{\bmr_{ij}} \rightarrow r_{\rm cut}$.

Hence, the basis functions have a nonlinear dependence on $\uc_{\mu_l}$. Therefore, we need to modify the active learning algorithm to take this nonlinearity into account. For this purpose, under the assumption that the vector $\umtheta$ now comprises the basis coefficients \emph{and} the $\uc_{\mu_l}$ parameters, we linearize \eqref{eq:fitting} with respect to some initial $\bar{\umtheta}$ to first order such that
\begin{equation}
 \Esite(\Atoms;\umtheta) - \Esite^{\rm qm}(\Atoms) \approx
 \Esite(\Atoms;\bar{\umtheta}) - \Esite^{\rm qm}(\Atoms) + \frac{\partial \Esite(\Atoms;\umtheta)}{\partial \umtheta}\bigg\vert_{\umtheta=\bar{\umtheta}} (\umtheta - \bar{\umtheta})
 = L(\umtheta) = 0.
\end{equation}
The system matrix associated with the new loss function $L$ is then the $n \times m$ Jacobian matrix
\begin{equation}
 \uuJ =
 \begin{pmatrix}
  \frac{\partial \Esite(\Neigh_1;\umtheta)}{\partial \mtheta_1}\Big\vert_{\umtheta=\bar{\umtheta}} & \cdots & \frac{\partial \Esite(\Neigh_1;\umtheta)}{\partial \mtheta_m}\Big\vert_{\umtheta=\bar{\umtheta}} \\
  \vdots & \ddots & \vdots \\
  \frac{\partial \Esite(\Neigh_n;\umtheta)}{\partial \mtheta_1}\Big\vert_{\umtheta=\bar{\umtheta}} & \cdots & \frac{\partial \Esite(\Neigh_n;\umtheta)}{\partial \mtheta_m}\Big\vert_{\umtheta=\bar{\umtheta}}
 \end{pmatrix},
\end{equation}
where $n$ is the total number of atomic neighborhoods in the training set which is usually significantly larger than the size of the parameter vector $m$.

We proceed as in Section \ref{sec:algo} by finding an $m \times m$ submatrix $\uuA$ of $\uuJ$ using the maxvol algorithm. The extrapolation grade $\gamma$ of a neighborhood $\AtomsNew$ is now defined---analogously to \eqref{eq:comp_extrap_grade}---as the maximum absolute element of the vector
\begin{equation}\label{eq:comp_extrap_grade_nln}
 \begin{aligned}
  \uc &= \ub^\sT\uuA^{-1} \\
  &=
  \begin{pmatrix}
   \frac{\partial \Esite(\NeighNew;\umtheta)}{\partial \mtheta_1}\Big\vert_{\umtheta=\bar{\umtheta}} & \cdots & \frac{\partial \Esite(\NeighNew;\umtheta)}{\partial \mtheta_m}\Big\vert_{\umtheta=\bar{\umtheta}}
  \end{pmatrix}
  \begin{pmatrix}
   \frac{\partial \Esite(\Neigh_1;\umtheta)}{\partial \mtheta_1}\Big\vert_{\umtheta=\bar{\umtheta}} & \cdots & \frac{\partial \Esite(\Neigh_1;\umtheta)}{\partial \mtheta_m}\Big\vert_{\umtheta=\bar{\umtheta}} \\
   \vdots & \ddots & \vdots \\
   \frac{\partial \Esite(\Neigh_m;\umtheta)}{\partial \mtheta_1}\Big\vert_{\umtheta=\bar{\umtheta}} & \cdots & \frac{\partial \Esite(\Neigh_m;\umtheta)}{\partial \mtheta_m}\Big\vert_{\umtheta=\bar{\umtheta}}
  \end{pmatrix}^{-1}.
 \end{aligned}
\end{equation}

\subsection{Training procedure}
\label{sec:mtp_train}

Since we perform structural relaxation and compute energy barriers, we need to fit the MTP with respect to energies and forces. Therefore, we compute the hyperparameters $\umtheta$ by minimizing the loss functional
\begin{equation}
 L(\umtheta) =
 C_\rme \sum_{\{ \Atom_i^{\rm tr} \} \in \scT} \left( \Etot(\{ \Atom_i^{\rm tr} \};\umtheta) - \Etot^{\rm qm}(\{ \Atom_i^{\rm tr} \}) \right) +
 C_\rmf \sum_{\{ \Atom_i^{\rm tr} \} \in \scT}\sum_{\Atom_i \in \{ \Atom_i^{\rm tr} \}} \left( \bforce_\atom(\{ \Atom_i^{\rm tr} \};\umtheta) - \bforce^{\rm qm}_\atom(\{ \Atom_i^{\rm tr} \}) \right) +
 C_\rmr (\umtheta^\sT \cdot \umtheta),
\end{equation}
where $\bforce_\atom$, $\bforce^{\rm qm}_\atom$ are the MTP, respectively, the quantum mechanical forces on an atom $\Atom_i$ corresponding to a specific configuration $\{ \Atom_i^{\rm tr} \}$ in the training set $\scT$. The $C$-constants are regularization parameters which we set to
\begin{align}
 C_\rme = 1, && C_\rmf = 0.01, && C_\rmr = 10^{-6},
\end{align}
throughout this work.

\section{Equivalence of the quadrupole- and dipole-like configurations}
\label{sec:dipole_config_analysis}

In the following we show that the quadrupole- and the dipole-like configurations, enclosed by the dashed and continuous lines in Figure \ref{fig:construct_dft_cells} (3), respectively, are equivalent when the displacement is the anisotropic linear elastic solution
\begin{equation}
 \bdisplDD(\bmx)
 = \frac{1}{2} \sum_{i=1}^3\ln{((x + p_iy)^2 + y^2)^2}\bmc_i - \arctantwo{(q_iy,x+p_iy)}\bmd_i
\end{equation}
of a screw dislocation. Assuming that the $\rmx$, $\rmy$ and $\rmz$-axes correspond to the $\langle 112\rangle$, $\langle 110 \rangle$ and $\langle 111\rangle$ direction, namely the slip direction, the slip plane normal and the dislocation line direction, the real quantities $p_i$, $q_i$, $\bmc_i$ and $\bmd_i$ can be computed for $\bburgers = \begin{pmatrix} 0 & 0 & \burgers \end{pmatrix}^\sT$ using the formalism described in \citep[][p. 444--445]{hirth_theory_1982}.

We now need to show that the displacement \eqref{eq:displ_mirr} on the upper boundary of the rectangular region with the $\AtomsNew$-atoms is the same as on the lower boundary of the rectangular region enclosing the $\{ \Atom_i^1 \}$-atoms, and vice versa. That is, $\forall\,x \in (-l_\rmx/2,l_\rmx/2] \; \wedge \; \forall\,x' \in (-3l_\rmx/2,-l_\rmx/2]$, we have $\bdispl(x,l_\rmy/2,z) = \bdispl(-x'-l_\rmx,-l_\rmy/2,z)$ or, equivalently, $\bdispl(x,l_\rmy/2,z) = \bdispl(-x,-l_\rmy/2,z)$. The equivalence is therefore obvious for the logarithmic part. For the arctan contribution we write $\arctantwo{(q_iy/(x+p_i))} = \arctantwo{(\bar{y}/\bar{x})}$. Assuming that $\bar{y}>0$, it follows for
\begin{equation}
 {\everymath={\displaystyle}
  \begin{array}{@{}r@{{}\mathrel{}}l@{\mathrel{}{}}l@{}}
   \bar{x} < 0: & \qquad
   \arctantwo{(\bar{y},\bar{x})} - \arctantwo{(-\bar{y},-\bar{x})}
   = \pi - 0 = \pi, \\
   \bar{x} = 0: & \qquad
   \arctantwo{(\bar{y},0)} - \arctantwo{(-\bar{y},0)}
   = \frac{\pi}{2} - \left(-\frac{\pi}{2}\right) = \pi, \\
   \bar{x} > 0: & \qquad
   \arctantwo{(\bar{y},\bar{x})} - \arctantwo{(-\bar{y},-\bar{x})}
   = 0 - (-\pi) = \pi.
  \end{array}
 }
\end{equation}
Thus, we have
\begin{equation}
 \displ_\rmz(x,l_\rmy/2,z) - \displ_\rmz(-x,-l_\rmy/2,z) = \frac{1}{2} \sum_{i=1}^3 \pi d_{3,i} = \frac{\burgers}{2}
\end{equation}
since $\sum_{i=1}^3 d_{3,i} = \burgers/\pi$ which can be deduced from the definition of the displacement at the branch cut. Therefore, the displacement field differs by the constant $\burgers/2$ which must be taken into account when constructing the training configurations. Hence, the region framed by the continuous line in Figure \ref{fig:construct_dft_cells} (3) and (4) must be the triclinic region
\begin{equation}\label{eq:triclinic_region}
 \left\{\, \bmx = \sum_{i=1}^3 \alpha_i \bmv_i \,\bigg\vert\, \alpha_i \in [0,1] \,\right\},
\end{equation}
spanned by the lattice vectors
\begin{align}
    \bmv_1 = \begin{pmatrix} 2l_\rmx & 0 & 0 \end{pmatrix}^\sT,
 && \bmv_1 = \begin{pmatrix} l_\rmx & l_\rmy & b/2 \end{pmatrix}^\sT,
 && \bmv_1 = \begin{pmatrix} 0 & 0 & b \end{pmatrix}^\sT.
\end{align}
The training configuration $\{ \Atom_i^{\rm tr} \}$ is then constructed by imposing the displacement \eqref{eq:displ_mirr} on the ideal lattice sites which are in \eqref{eq:triclinic_region}.

If the actual atomistic displacements follow a similar pattern, the training configurations can be constructed analogously. This argument trivially applies if the atomistic solution does not differ too much from linear elasticity outside the dislocation core---which is usually the case for compact core structures.

\section{DFT simulations}
\label{sec:VASP}

The parameters we have used in all our VASP calculations of bcc W are given in Table \ref{tab:VASP}. Electronic relaxation was performed using the preconditioned minimal residual method, as implemented in VASP, and terminated when the energy difference between two subsequent iterations was less than 10$^{-4}$\,eV.

With these parameters, we obtained the lattice constant $\latConst=3.168\,\text{\AA}$ and the cubic elastic constants $C_{11}=551.4\,\text{GPa}$, $C_{12}=200\,\text{GPa}$ and $C_{44}=139.9\,\text{GPa}$, using the \texttt{get\_elastic\_constants} function from Matscipy \citep{Matscipy}.

\begin{table}[hbt]
 \centering
 \begin{tabular}{r|c}
  Exchange-correlation & PE generalized gradient approximation \citep{perdew_generalized_1996} \\ \hline
  PAW potential & PAW\_PBE W 08Apr2002 \\ \hline
  Energy cut-off & 334.585\,eV \\ \hline
  $\bmk$-point mesh & 2$\times$3$\times$16 \\ \hline
  Smearing width & 0.06\,eV
 \end{tabular}
 \caption{VASP parameters}
 \label{tab:VASP}
\end{table}

\section{Algorithmic results and training errors for the atomistic simulations using MTPs}
\label{sec:train_err_mtp-eamX}

\subsection{Core relaxation}

\begin{table}[H]
 \renewcommand{\arraystretch}{1.2}
 \centering
 \begin{tabular}{c||c|c||c|c}
  & \multicolumn{2}{c||}{periodic} & \multicolumn{2}{c}{cylinder}\\
  & MTP-DFT & MTP-EAM4 & MTP-DFT & MTP-EAM4 \\
  \hline\hline
  iterations & 57 & 91 & 63 & 72 \\ \hline
  training set size & 44 & 35 & 40 & 55
 \end{tabular}
 \caption{Total number of iterations and training set size for the core relaxation problems in Section \ref{sec:cyl.relax}}
 \label{tab:screw_relax_iter_&_ts_size}
\end{table}

\begin{table}[H]
 \renewcommand{\arraystretch}{1.2}
 \centering
 \begin{tabular}{c||c|c|c}
  & \multicolumn{3}{c}{Error in:} \\
  & Energy [eV] & Energy per atom [eV] & Forces [eV/\AA] \\
  \hline\hline
  Maximal absolute difference & 0.00642 & 4.759$\cdot$10$^{-5}$ & 0.0476 \\ \hline
  Average absolute difference & 7.533$\cdot$10$^{-4}$ & 5.58$\cdot$10$^{-6}$ & 0.0089 \\ \hline
  RMS                         & 0.00115 & 8.509$\cdot$10$^{-6}$ & 0.0105 \\ \hline\hline
  Max(ForceDiff) / Max(Force) & & & 0.0164 \\ \hline
  RMS(ForceDiff) / RMS(Force) & & & 0.049
 \end{tabular}
 \caption{Training errors corresponding to the core relaxation problems for the cylindrical problem with the MTP-EAM4}
 \label{tab:screw_cyl_training_err_cyl_mtp-eam4}
\end{table}

\begin{table}[H]
 \renewcommand{\arraystretch}{1.2}
 \centering
 \begin{tabular}{c||c|c|c}
  & \multicolumn{3}{c}{Error in:} \\
  & Energy [eV] & Energy per atom [eV] & Forces [eV/\AA] \\
  \hline\hline
  Maximal absolute difference & 0.00293 & 2.167$\cdot$10$^{-5}$ & 0.0299 \\ \hline
  Average absolute difference & 5.258$\cdot$10$^{-4}$ & 3.895$\cdot$10$^{-6}$ & 0.0111 \\ \hline
  RMS                         & 7.457$\cdot$10$^{-4}$ & 5.524$\cdot$10$^{-6}$ & 0.0123 \\ \hline\hline
  Max(ForceDiff) / Max(Force) & & & 0.0378 \\ \hline
  RMS(ForceDiff) / RMS(Force) & & & 0.128
 \end{tabular}
 \caption{Training errors corresponding to the core relaxation problems for the periodic problem with the MTP-DFT}
 \label{tab:screw_cyl_training_err_per_mtp-dft}
\end{table}

\begin{table}[H]
 \renewcommand{\arraystretch}{1.2}
 \centering
 \begin{tabular}{c||c|c|c}
  & \multicolumn{3}{c}{Error in:} \\
  & Energy [eV] & Energy per atom [eV] & Forces [eV/\AA] \\
  \hline\hline
  Maximal absolute difference & 0.00681 & 5.048$\cdot$10$^{-5}$ & 0.0671 \\ \hline
  Average absolute difference & 0.00125 & 9.287$\cdot$10$^{-6}$ & 0.00699 \\ \hline
  RMS                         & 0.00169 & 1.254$\cdot$10$^{-5}$ & 0.00835 \\ \hline\hline
  Max(ForceDiff) / Max(Force) & & & 0.023 \\ \hline
  RMS(ForceDiff) / RMS(Force) & & & 0.0325
 \end{tabular}
 \caption{Training errors corresponding to the core relaxation problems for the periodic problem with the MTP-EAM4}
 \label{tab:screw_cyl_training_err_per_mtp-eam4}
\end{table}

\subsection{Peierls barrier}
\label{sec:training_errors_neb}

\begin{table}[H]
 \renewcommand{\arraystretch}{1.2}
 \centering
 \begin{tabular}{c||c|c||c|c}
  & \multicolumn{2}{c||}{periodic} & \multicolumn{2}{c}{cylinder}\\
  & MTP-DFT & MTP-EAM4 & MTP-DFT & MTP-EAM4 \\
  \hline\hline
  iterations & 32 & 57 & 38 & 61 \\ \hline
  training set size & 48 & 71 & 22 & 30
 \end{tabular}
 \caption{Total number of iterations and training set size for the Peierls barrier calculations in Section \ref{sec:cyl.neb}}
 \label{tab:screw_neb_iter_&_ts_size}
\end{table}

\begin{table}[H]
 \renewcommand{\arraystretch}{1.2}
 \centering
 \begin{tabular}{c||c|c|c}
  & \multicolumn{3}{c}{Error in:} \\
  & Energy [eV] & Energy per atom [eV] & Forces [eV/\AA] \\
  \hline\hline
  Maximal absolute difference & 0.0179 & 1.329$\cdot$10$^{-4}$ & 0.143 \\ \hline
  Average absolute difference & 0.0104 & 7.697$\cdot$10$^{-5}$ & 0.0206 \\ \hline
  RMS                         & 0.0122 & 9.073$\cdot$10$^{-5}$ & 0.0257 \\ \hline\hline
  Max(ForceDiff) / Max(Force) & & & 0.0448 \\ \hline
  RMS(ForceDiff) / RMS(Force) & & & 0.0699
 \end{tabular}
 \caption{Training errors corresponding to the NEB calculations for the cylindrical problem with the MTP-EAM4}
 \label{tab:screw_cyl_neb_training_err_cyl_mtp-eam4}
\end{table}

\begin{table}[H]
 \renewcommand{\arraystretch}{1.2}
 \centering
 \begin{tabular}{c||c|c|c}
  & \multicolumn{3}{c}{Error in:} \\
  & Energy [eV] & Energy per atom [eV] & Forces [eV/\AA] \\
  \hline\hline
  Maximal absolute difference & 0.00494 & 3.662$\cdot$10$^{-5}$ & 0.0721 \\ \hline
  Average absolute difference & 0.00215 & 1.593$\cdot$10$^{-5}$ & 0.0125 \\ \hline
  RMS                         & 0.00259 & 1.919$\cdot$10$^{-5}$ & 0.0139 \\ \hline\hline
  Max(ForceDiff) / Max(Force) & & & 0.0786 \\ \hline
  RMS(ForceDiff) / RMS(Force) & & & 0.191
 \end{tabular}
 \caption{Training errors corresponding to the NEB calculations for the periodic problem with the MTP-DFT}
 \label{tab:screw_cyl_neb_training_err_per_mtp-dft}
\end{table}

\begin{table}[H]
 \renewcommand{\arraystretch}{1.2}
 \centering
 \begin{tabular}{c||c|c|c}
  & \multicolumn{3}{c}{Error in:} \\
  & Energy [eV] & Energy per atom [eV] & Forces [eV/\AA] \\
  \hline\hline
  Maximal absolute difference & 0.0192  & 1.423$\cdot$10$^{-4}$ & 0.127 \\ \hline
  Average absolute difference & 0.00407 & 3.017$\cdot$10$^{-5}$ & 0.0145 \\ \hline
  RMS                         & 0.00522 & 3.866$\cdot$10$^{-5}$ & 0.0204 \\ \hline\hline
  Max(ForceDiff) / Max(Force) & & & 0.0325 \\ \hline
  RMS(ForceDiff) / RMS(Force) & & & 0.0857
 \end{tabular}
 \caption{Training errors corresponding to the NEB calculations for the periodic problem with the MTP-EAM4}
 \label{tab:screw_cyl_neb_training_err_per_mtp-eam4}
\end{table}

\subsection{Peierls stress}

\begin{table}[H]
 \renewcommand{\arraystretch}{1.2}
 \centering
 \begin{tabular}{c||c|c|c}
  & \multicolumn{3}{c}{Error in:} \\
  & Energy [eV] & Energy per atom [eV] & Forces [eV/\AA] \\
  \hline\hline
  Maximal absolute difference & 0.0167 & 1.238$\cdot$10$^{-4}$ & 0.249 \\ \hline
  Average absolute difference & 0.0058 & 4.295$\cdot$10$^{-5}$ & 0.0157 \\ \hline
  RMS                         & 0.00683 & 5.06$\cdot$10$^{-5}$ & 0.0201 \\ \hline\hline
  Max(ForceDiff) / Max(Force) & & & 0.0704 \\ \hline
  RMS(ForceDiff) / RMS(Force) & & & 0.0456
 \end{tabular}
 \caption{Training errors corresponding to the Peierls stress calculation for the MTP-EAM3 in Section \ref{sec:cyl.peierls_stress}. The simulation was carried out until the dislocation was about to pass the second Peierls valley. At this point the training set contained 51 configurations}
 \label{tab:screw_cyl_strain_training_err_eam3}
\end{table}

\end{appendix}


\AtEndEnvironment{thebibliography}{
 \bibitem{Matscipy}Matscipy, \url{https://github.com/libAtoms/matscipy}
}
\nocite{stukowski_visualization_2010}

\section*{References}
\bibliographystyle{elsarticle-harv}
\bibliography{references}

\end{document}

%% file: style.tex
\newcommand{\sty}[1]{\boldsymbol{#1}}

\newcommand{\ubar}[1]{\mkern 0.5mu\underline{\mkern-0.5mu#1\mkern-0.5mu}\mkern 0.5mu}
\newcommand{\uubar}[1]{\ubar{\ubar{#1}}}

\let\epsilon\varepsilon
\let\mtheta\theta
\let\theta\vartheta
\let\mrho\rho
\let\rho\varrho
\let\mphi\phi
\let\phi\varphi
\let\Gamma\varGamma
\let\Delta\varDelta
\let\Theta\varTheta
\let\Lambda\varLambda
\let\Xi\varXi
\let\Pi\varPi
\let\Sigma\varSigma
\let\Upsilon\varUpsilon
\let\Phi\varPhi
\let\Psi\varPsi

\let\Omega\varOmega

\newcommand{\abs}[1]{\vert #1 \vert}

\renewcommand{\det}[1]{\mathrm{det}\left(#1\right)} 

\newcommand{\bmc}{\sty{c}}
\newcommand{\bmd}{\sty{d}}

\newcommand{\bmk}{\sty{k}}

\newcommand{\bmr}{\sty{r}}

\newcommand{\bmv}{\sty{v}}

\newcommand{\bmx}{\sty{x}}

\newcommand{\clE}{\mathcal{E}}

\newcommand{\sT}{\mathsf{T}} 

\newcommand{\scT}{\mathscr{T}}

\newcommand{\rme}{\mathrm{e}}
\newcommand{\rmf}{\mathrm{f}}

\newcommand{\rmp}{\mathrm{p}}

\newcommand{\rmr}{\mathrm{r}}

\newcommand{\rmx}{\mathrm{x}}
\newcommand{\rmy}{\mathrm{y}}
\newcommand{\rmz}{\mathrm{z}}

\newcommand{\ub}{\ubar{b}}
\newcommand{\uc}{\ubar{c}}

\newcommand{\umtheta}{\ubar{\mtheta}}

\newcommand{\uuA}{\uubar{A}}

\newcommand{\uuJ}{\uubar{J}}
